\documentclass[traditabstract]{aa}
\usepackage{graphicx}
\usepackage{rotating,subfigure,amssymb,afterpage}
\usepackage{txfonts}
\usepackage{natbib}
\usepackage[pdftitle={{\sf REFLEX II} Cluster X-ray Luminosity Function},
pdfauthor={H. B\"ohringer et al.},
pdfsubject={}, pdfkeywords={}, bookmarks, bookmarksopen, 
colorlinks=true, linkcolor=blue, citecolor=blue, anchorcolor=blue,
urlcolor=blue]{hyperref} 
\newcommand{\propsim}{\lower 3pt \hbox{$\, \buildrel {\textstyle
      \propto}\over {\textstyle \sim}\,$}}
\begin{document}
   \title{The extended ROSAT-ESO Flux Limited X-ray Galaxy Cluster
   Survey (REFLEX II)\\ IV. X-ray Luminosity Function and First Constraints 
        on Cosmological Parameters
\thanks{
   Based on observations at the European Southern Observatory La Silla,
   Chile}}

   \author{Hans B\"ohringer\inst{1}, Gayoung Chon\inst{1}, Chris A. Collins\inst{2}}

   \offprints{H. B\"ohringer, hxb@mpe.mpg.de}

   \institute{$^1$ Max-Planck-Institut f\"ur extraterrestrische Physik,
                   D-85748 Garching, Germany.\\
              $^2$ Astrophysics Research Institute, Liverpool John Moores University, 
                    IC2, Liverpool Science Park, 146 Brownlow Hill,
                    Liverpool L3 5RF, UK}

   \date{Submitted 29/11/13}

\abstract{
The X-ray luminosity function which is closely related to the cluster mass function
is an important statistic of the census of galaxy clusters in our Universe. It is also
an important means to probe the cosmological model of our Universe.
Based on our recently completed {\sf REFLEX II} cluster sample comprising 910 galaxy clusters 
with redshifts we construct the X-ray luminosity function of galaxy clusters 
for the nearby Universe and discuss its implications.
We derive the X-ray luminosity function of the {\sf REFLEX II} clusters on the basis of
a precisely constructed selection function for the full sample and for several redshift
slices from $z = 0$ to $z = 0.4$. In this redshift interval we find no significant
signature of redshift evolution of the luminosity function. 
We provide the results of fits of a parameterized Schechter function and extensions
of it which provide a reasonable characterization of the data.
We also use a model for structure formation and galaxy cluster evolution to compare
the observed X-ray luminosity function with the theoretical predictions for
different cosmological models. The most interesting constraints can be derived
for the cosmological parameters $\Omega_m$ and $\sigma_8$. 
We explore the influence of several model
assumptions on which our analysis is based. We find that the scaling relation
of X-ray luminosity and mass introduces the largest systematic uncertainty.
From the statistical uncertainty alone we can constrain the matter density 
parameter, $\Omega_m \sim 0.27 \pm 0.03$ and the amplitude parameter of the 
matter density fluctuations, $\sigma_8 \sim 0.80 \pm 0.03$. Marginalizing over
the most important uncertainties, the normalisation and slope of the
$L_X - M$ scaling relation, we have larger error bars and a result of
$\Omega_m \sim 0.29 \pm 0.04$  and $\sigma_8 \sim 0.77 \pm 0.07$ ($1\sigma$
confidence limits).
We compare our results with those of the SZ-cluster survey
provided by the {\sf PLANCK} mission and we find very good agreement with the results
using {\sf PLANCK} clusters as cosmological probes, but we have some tension 
with {\sf PLANCK} cosmological results from the microwave background anisotropies,
which we discuss in the paper. We also make a comparison with
results from the SDSS cluster survey, several cosmological X-ray cluster surveys,
and recent Sunyaev-Zel'dovich effect surveys. We find good agreement with 
these previous results and show that the {\sf REFLEX II} survey provides a significant 
reduction in the uncertainties compared to earlier measurements.}

\authorrunning{B\"ohringer et al.}
\titlerunning{{\sf REFLEX II} Cluster X-ray Luminosity Function}
   \maketitle
%

\section{Introduction}

Galaxy clusters as the largest, clearly defined objects in
our Universe are interesting astrophysical laboratories 
and important cosmological probes (e.g. Sarazin 1986, Borgani et al. 2001,
Voit 2005, Vikhlinin et al. 2009, Allen et al. 2011, B\"ohringer 2011).
They are particularly good tracers of the large-scale structure
of the cosmic matter distribution and its growth with time.
While most of the precise knowledge on the galaxy cluster population
has come from X-ray observations as detailed in the above references, 
recent progress has also been made by optical
cluster surveys (e.g. Rozo et al. 2010) and millimeter wave surveys
using the Sunyaev-Zel'dovich effect 
(Reichardt et al. 2012, Benson et al. 2013, Marriage et al. 2011, 
Sehgal et al. 2011, PLANCK-Collaboration 2011, 2013b). 
X-ray surveys for galaxy clusters are still most advanced providing  
statistically well defined, approximately mass selected
cluster samples, since: (i) X-ray luminosity is tightly correlated 
to mass (e.g. Reiprich \& B\"ohringer 2002, Pratt et al. 2009), (ii) bright
X-ray emission is only observed for evolved clusters with
deep gravitational potentials, (iii) the X-ray emission is
highly peaked and projection effects are minimized,
and (iv) for all these reasons the survey selection function
can be accurately modeled.

The {\sf ROSAT} All-Sky Survey (RASS, Tr\"umper 1993) is the only 
existing full sky survey conducted with
an imaging X-ray telescope, providing a sky atlas in which one can
search systematically for clusters in the nearby Universe.
The largest, high-quality sample of X-ray selected galaxy
clusters is provided so far by the {\sf REFLEX} Cluster Survey 
(B\"ohringer et al. 2001, 2004, 2013) based on the southern 
extragalatic sky of RASS at declination $\le 2.5$ degree.  
The quality of the sample has been demonstrated by showing
that it can provide reliable measures of the large-scale structure
(Collins et al. 2000, Schuecker et
al. 2001a, Kerscher et al. 2001), yielding cosmological parameters 
(Schuecker et al.  2003a, b; B\"ohringer 2011) in good agreement 
within the measurement uncertainties with the subsequently published
WMAP results (Spergel et al. 2003, Komatsu et al. 2011).
The {\sf REFLEX} data have also been used to study the X-ray
luminosity function of galaxy clusters (B\"ohringer et al. 2002),
the galaxy velocity dispersion - X-ray luminosity relation 
(Ortiz-Gil et al., 2004), the statistics of
Minkowski functionals in the cluster distribution (Kerscher et al. 2001),
and to select statistically well defined subsamples like 
the HIFLUGCS (Reiprich \& B\"ohringer 2002) and {\sf REXCESS} (B\"ohringer
et al. 2007). The latter is particularly important as a representative
sample of X-ray surveys to establish X-ray scaling relations 
(Croston et al. 2008, Pratt et al. 2009, 2010, Arnaud et al. 2010) 
and the statistics of the morphological
distribution of galaxy clusters in X-rays (B\"ohringer et al. 2010).

Recently we have completed an extension of {\sf REFLEX} apart from 11
missing redshifts, {\sf REFLEX II}, which about doubles the size of
the cluster sample. The construction of this sample is described in
B\"ohringer et al. (2013). In the present paper we describe the construction 
of the {\sf REFLEX II} X-ray luminosity function from the galaxy cluster 
data and the survey selection function derived in B\"ohringer et al. 
(2013). We fit parameterized functions to the data and we compare the
observed luminosity function to the predictions of cosmological structure 
formation models. From the latter comparison we obtain constraints 
on cosmological parameters. The most sensitive of these parameters 
are the matter density parameter, $\Omega_m$, and the amplitude 
parameter of the matter density fluctuations, $\sigma_8$. We focus on
the derivation of robust constraints on the two parameters in this 
paper, while leaving a comprehensive modeling of the combined 
uncertainty of all relevant cosmological and cluster parameters 
to a future publication.  We study the errors introduced by various 
other uncertainties rather case by case and evaluate the overall 
systematic errors. The {\sf REFLEX II} cluster sample has also 
recently been used to construct the first supercluster catalog for
clusters with a well defined selection function (Chon \& B\"ohringer 
2013), showing among other results that the X-ray luminosity
function of clusters in superclusters is top-heavy in 
comparison to that of clusters in the field.

A preliminary sample of {\sf REFLEX II} which had 49 redshifts less than
used here, has been applied to the study of the galaxy cluster power
spectrum by Balaguera-Antolinez et al. (2011). The results show a very
good agreement with the cosmological predictions based on cosmological 
parameters determined from WMAP 5 yr data. In a second paper
(Balaguera-Antolinez et al. 2012), 
in which the construction of {\sf REFLEX} mock samples from simulations 
used in the earlier paper is described, 
a preliminary X-ray luminosity function of 
{\sf REFLEX II} has been determined. Here we use a completely new
approach with updates on the cluster sample, the scaling relations, and
the missing flux correction used in the sample construction, and the 
survey selection function based on the procedures described in 
B\"ohringer et al. (2013). 

Other previous determinations of the X-ray luminosity function of galaxy
clusters include: Piccinotti et al. (1982), Kowalski et al. (1984), Gioia
et al. (1984), Edge et al. (1990), Henry et al. (1992), Burns et al. (1996),
Ebeling et al. (1997), Collins et al. (1997), Burke et al. (1997),
Rosati et al. (1998), Vikhlinin et al. (1998), De Grandi et al. (1999),
Ledlow et al. (1999), Nichol et al. (1999), Gioia et al (2001), 
Donahue et al. (2001) Allen et al. (2003), Mullis et al. (2004), 
B\"ohringer et al. (2007), Koens et al. (2013). 

The paper is organized as follows. In chapter 2 we introduce the REFLEX II
galaxy cluster sample and the survey selection function. In section 3 
we use the parameterized Schechter function fitted to our data to 
describe the resulting X-ray luminosity function. In section 4 we 
outline the cosmological modeling used for the theoretical prediction 
of the cluster mass and X-ray luminosity function. In section 5 we 
discuss the results of the model comparison to the data for different 
cosmological models. The effect of the uncertainties in the used
cluster scaling relations on the results is discussed in section 
6 and other systematic uncertainties of our analysis are 
discussed in section 7. In section 8 we compare our results to findings
from other surveys and section 9 closes the paper with the summary 
and conclusions.

If not stated otherwise, we use for the calculation of physical 
parameters and survey volumes a geometrically flat
$\Lambda$-cosmological model with $\Omega_m = 0.3$ and $h_{70} = H_0/70$ km s$^{-1}$ 
Mpc$^{-1}$ = 0.7. All uncertainties without further 
specifications refer to 1$\sigma$ confidence limits.

\section{The REFLEX II Galaxy Cluster Survey}

The REFLEX II galaxy cluster survey is based on the detection of
galaxy clusters in the RASS (Voges et al. 1999). The region of the survey is
the southern sky below equatorial latitude +2.5 deg. at galactic
latitude $b_{II} \ge 20$ deg. The regions of the Magellanic clouds have
been masked. The survey region selection, the source detection, the
galaxy cluster sample definition and compilation, and the construction of
the survey selection function  as well as tests of the completeness of the
survey are described in B\"ohringer et al. (2013). In brief the survey area
is $ \sim 2.4$ ster or 13924 square degrees. The nominal flux limit down to which
galaxy clusters have been identified in the RASS in this region is
$1.8 \times 10^{-12}$ erg s$^{-1}$ cm$^{-2}$ in the
0.1 - 2.4 keV energy band. For the construction of the
X-ray luminosity function in this paper we impose an additional cut
on the minimum number of detected source photons of 20 counts. This has
the effect that the nominal flux cut quoted above is only reached in about
80\% of the survey and in regions with lower exposure and higher interstellar
absorption the flux limit is accordingly higher 
(see Fig.\ 11 in B\"ohringer et al. 2013). This effect is modeled and
taken into account in the survey selection function.

The flux limit imposed on the survey is for a nominal flux, that has been
calculated from the detected photon count rate for a cluster X-ray spectrum
characterized by a temperature of 5 keV, a metallicity of 0.3 solar,
a redshift of zero, and  an interstellar absorption column
density given by the 21cm sky survey described
by Dickey and Lockmann (1990). This count rate to flux conversion is
appropriate prior to any redshift information and is analogous to an observed
object magnitude corrected for galactic extinction in the optical.

After the redshifts have been measured, a new flux is calculated taking the
redshifted spectrum and an estimate for the spectral temperature
into account. The temperature estimate is obtained from the X-ray luminosity -
temperature relation of Pratt et al. (2009) determined from the {\sf REXCESS} cluster
sample, which is a sample of clusters drawn from {\sf REFLEX I} for deeper follow-up
observations with XMM-Newton and which is representative of the entire flux limited
survey (B\"ohringer et al. 2007). The luminosity is determined first from the
observed flux by means of the luminosity distance for a given redshift. Using
the X-ray luminosity mass relation given in Pratt et al. (2009) we can then
use the mass estimate to determine a fiducial radius of the cluster, which is
taken to be $r_{500}$ \footnote{$r_{500}$ is the radius where the average
mass density inside reaches a value of 500 times the critical density
of the Universe at the epoch of observation.}. We then use a beta model for the
cluster surface brightness distribution to correct for the possibly missing
flux in the region between the detection aperture of the source photons and
the radius $r_{500}$. The procedure to determine the flux, the luminosity,
the temperature estimate, and $r_{500}$ is done iteratively and described in
detail in B\"ohringer et al. (2013). In that paper we deduced a mean flux 
uncertainty for the {\sf REFLEX II} clusters of 20.6\%, which is 
mostly due to the Poisson statistics of the source counts but also 
contains some systematic errors. In this paper we will for the following
analysis adopt a flux and luminosity measurement uncertainty of 
20\%.

The X-ray source detection and selection is based on the official RASS source
catalogue by Voges et al. (1999). We have been using the publically available
final source catalog \footnote{the RASS source catalogs can
be found at: http://www.xray.mpe.mpg.de/rosat/survey/rass-bsc/  for the
bright sources and http://www.xray.mpe.mpg.de/rosat/survey/rass-fsc/ for
the faint sources} as well as a preliminary source list that
was created in the course of the production of the public catalog.
The latter source list had a lower significance limit and included
a larger number of detections, but it had not been 
manually screened as the public data set. This ensured a higher completeness
of the input catalog and spurious sources are caught by our subsequent
screening.
Since the analysis software used to produce the public RASS source catalog
is tuned for point sources and does not perform so well for the extended
sources of galaxy clusters, we have reanalyzed all the X-ray sources with the
growth curve analysis method (B\"ohringer et al. 2000). The flux cut was imposed
on the reanalysed data set. The results of the flux determination was inspected
visually for all sources above and near the flux limit and pathological cases
as well as cases of source confusion have been corrected manually.

The galaxy clusters among the sources have been identified using all available
means: X-ray source properties, available optical images
(mostly from DSS\footnote{see http://archive.stsci.edu/dss/}), 
literature data (mostly from NED\footnote{see http://ned.ipac.caltech.edu/})
and finally by follow-up observations at ESO La Silla. The source identification
scheme is described in detail in B\"ohringer et al. (2013). The redshifts have
been secured mostly by multi-object spectroscopy and the redshift accuracy
of the clusters is typically 60 km/s (Guzzo et al. 2009, 
Chon \& B\"ohringer 2012).

\begin{figure}
   \includegraphics[width=\columnwidth]{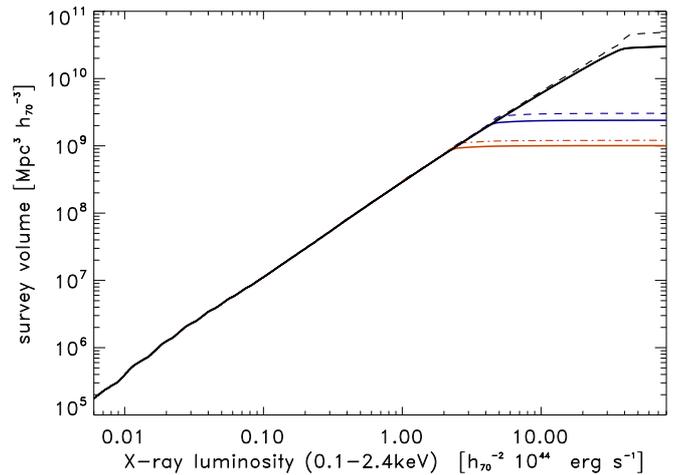}
\caption{Effective survey volume as a function of X-ray luminosity.
The survey volume has been calculated for three different cut-off redshift,
$z=0.8$, $z=0.3$, $z=0.22$ for our reference cosmology
($h=0.7$, $\Omega_m = 0.3$ and  $\Omega_{\Lambda}=0.7$). We have also determined
the same set of curves for a cosmology with the parameters
$h=0.7$, $\Omega_m = 0.26$ and $\Omega_{\Lambda}=0.74$
shown as dashed curves. {\bf Note} that the difference
between the set of curves has been amplified by a factor of 10, to make 
the offset better visible. This affects the dashed curves which are shown 
slightly offset from the original position.
}\label{fig1}
\end{figure}

The survey selection function is a very important survey product that is crucial
for the work of this paper. We have constructed the selection function in the
form of a survey mask that provides the limiting X-ray luminosity for the cluster
detection as a function of the sky position and redshift. For the sky position
the survey mask is currently given in pixels of one square degree. The survey
mask takes all the systematics of the RASS exposure distribution, galactic
absorption, the fiducial flux, the detection count limit, and all the applied 
corrections described above into account. The survey mask is given in three dimensional
form so that it can also be used for any study that is related to the spatial
distribution of the clusters. A preliminary version of it has been used for the
construction of the cluster density distribution power spectrum and a preliminary
construction of the luminosity function in Balaguera-Antolinez et al. (2011, 2012).
It allows further to select a cluster sample from simulations in a 
precisely analogous way as the {\sf REFLEX II} survey selects the clusters from the sky.

The survey selection function provides the means to calculate the effective survey
volume as a function of the X-ray luminosity. This survey volume function is shown in
Fig.~\ref{fig1} for different imposed redshift limits. The objects with
the lowest X-ray luminosities are only detected in a small volume in the nearby
Universe due to the flux limit of the survey. The luminous clusters with
$L_X \ge 2.5 \times 10^{44}$ erg s$^{-1}$ are found in a volume larger than 1 Gpc$^3$.
We also show the survey volume calculated for two different
cosmological models in the Figure including the reference cosmology
used in the paper ( $h=0.7$, $\Omega_m = 0.3$ and  $\Omega_{\Lambda}=0.7$)
and a cosmology closer the WMAP results (Komatsu et al. 2011) with
parameter values of $h=0.7$, $\Omega_m = 0.26$ and $\Omega_{\Lambda}=0.74$.
The difference is rather small and we have exaggerated the difference between the
curves representing the two cosmologies by a factor of 10 to make it better
visible. The largest difference is found for 
the survey volume at the cut-off redshift,
since at lower redshifts the effect of the cosmological model on the deduced
luminosity and survey volume partly compensate to produce very similar curves.

In the following we will use different versions of these calculations, 
to also determine the selection function in redshift shells. For the 
proper cosmological modeling of the results we determine the survey volume
for any given cosmology used in the model fitting process.  

\section{The X-ray luminosity function}

\begin{figure}
   \includegraphics[width=\columnwidth]{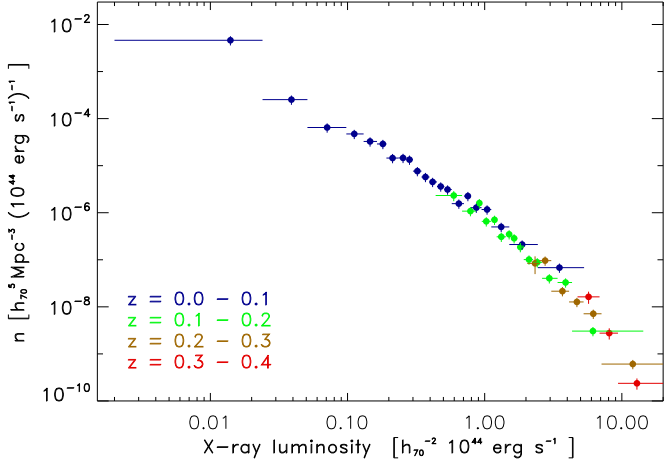}
\caption{X-ray luminosity function of {\sf REFLEX II} determined
in four equidistant redshift shells from $z = 0$ to $z = 0.4$.
Due to the flux limit of the sample the different redshift 
shells cover different luminosity ranges. In the overlap region
the functions show no major differences.
}\label{fig2}
\end{figure}

\begin{figure}
   \includegraphics[width=\columnwidth]{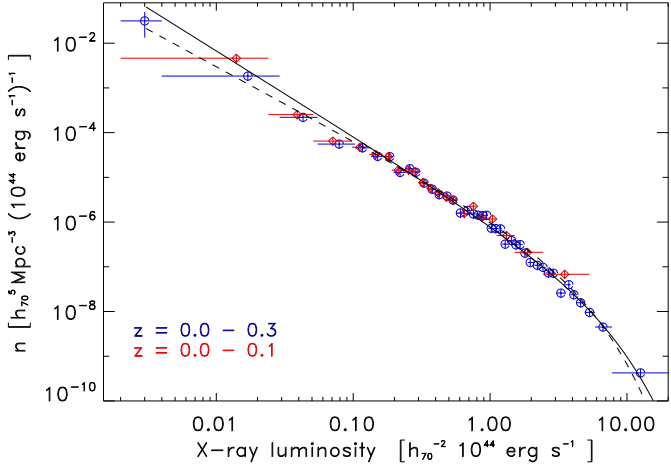}
\caption{X-ray luminosity function of the {\sf REFLEX II} galaxy
cluster survey  averaged over the survey volumes out to redshift
$z = 0.1$ and $z = 0.3$. Schechter functions have been fit to the
two data sets separately to visualize the potential difference of the 
functions. The fits were performed over the whole observed X-ray
luminosity range.
}\label{fig3}
\end{figure}

The X-ray luminosity function (XLF) is determined from the 
catalog of clusters and the survey selection function in the 
form of the effective survey volume as a function of X-ray luminosity 
as shown in Fig.~\ref{fig1} . We use a source detection count cut of 
minimum 20 photons for the selection. Then the binned differential
X-ray luminosity function is given by

\begin{equation}
{dn(L_X) \over dL_X} = {1 \over \Delta L_X} \sum_i {1 \over V_{max}(L_{X_i})}
\end{equation}

where $V_{max}$ is the effective detection volume and $\Delta L_X$ is 
the width of the luminosity bin and the sum includes
all clusters in the bin. Figs.~\ref{fig2} and ~\ref{fig3} show 
the XLF derived for a binning with 20 
clusters per bin, except for the bin at the lowest X-ray luminosity,
for different redshift ranges. While Fig.~\ref{fig2} presents
the XLF in four equidistant redshift shells from $z = 0$ to $z = 0.4$,

The luminosity values used in the construction of the XLF
are the luminosities inside $r_{500}$, corrected for missing flux
in the 0.1 - 2.4 keV rest frame band.
The errors for the XLF given in the Figures are the Poisson
uncertainties for the number of clusters per bin. 

Looking for evolutionary effects in the XLF in 
Fig.~\ref{fig2}, we do not
detect any significant difference of the functions, which implies
no severe deficiencies in the cluster detection
in the high redshift shells and no strong
evolutionary effects. What is expected
from theory is mostly a change at the high luminosity end, where
the lower redshift shells should show more luminous clusters
than the more distant ones. In Appendix A we show the expectation
for our best fitting cosmological model. While the mass function
would show a more noticeable change with redshift, the corresponding change 
in the XLF is small as the evolution effect is
partly compensated by the adopted
redshift evolution of the X-ray luminosity - mass relation. 
In the theoretical functions we see a significant difference
only at luminosities above about $6 \times 10^{44}~h_{70}^{-2}$ erg
s$^{-1}$ where we have hardly any objects and no statistics in the
two lowest redshift shells below $z = 0.2$. 

Fig.~\ref{fig3} compares the XLF in the two redshift ranges from
z = 0 to $z_{max} = 0.1$ and $z_{max} = 0.3$, respectively. There are 
419 clusters at $z < 0.1$ and 802 clusters at $z < 0.3$ for 
$L_X \ge 3 \times 10^{42}$ erg s$^{-1}$.
In the larger volume at $z > 0.3$
there are 53 additional clusters. Constructing the 
XLF in an even larger volume (e.g. for $z_{max} = 0.8$) shows hardly any 
difference in the resulting function. To make it
even more clear that there is not much leverage to look for redshift
evolution in the data, we fit Schechter 
functions to the two data sets (as will be explained in the next section),
which are also shown in the Figure.
The Schechter fit prefers a slightly less top-heavy luminosity function
for the lowest redshift bin. We will therefore not further 
pursue a detailed modeling of the evolution of the XLF 
in this paper and assume that the XLF can be described reasonably 
well by a constant function in the redshift range $z = 0 - 0.4$.

\subsection{Fits of a Schechter function}

For an analytical, phenomenological description of the 
{\sf REFLEX} X-ray luminosity function we fit a 
Schechter function of the form

\begin{equation}
{n(L_X)~dL_X}~ =~ n_0~ \left( {L_X \over L_X^{\ast}}\right)^{-\alpha}
exp\left(- {L_X \over L_X^{\ast}}\right)  {dL_X \over L_X^{\ast}}
\end{equation}

\noindent
to the data. 

We use a maximum likelihood method to determine the 
best fitting Schechter function parameters by comparing the 
predicted and observed X-ray luminosity distribution of the 
galaxy clusters. The approach we take is similar to what we used in
Schuecker \& B\"ohringer (1998) and B\"ohringer et al. (2002; 
see also Daley \& Vere-Jones 1988 and for a similar application 
Marshall et al. 1983 and Henry 2004).
To test a distribution function, $\lambda (x)$, with discrete 
observational data points, $\lambda (x_i)$, we minimize the likelihood
function:

\begin{equation}
ln~ L = - \int \lambda (x) dx + \sum_{i=1}^{N}~
ln \lambda (x_i) ~~~~.  
\end{equation}

The distribution function $\lambda (x)$ should describe observables.
In our case we compare the observed and predicted X-ray luminosity
distribution of the clusters. The X-ray luminosities can be considered
to be almost direct observables, as the spectral model 
assumptions used for their derivation are safe and possible changes 
have very small influence on the resulting luminosities.
Before the comparison, the predicted X-ray luminosity
distribution, $N(L_X)$, is folded with the observational 
error in the following way:

\begin{equation}
N(L_X) ~ =~ \int_{L_{X min}}^{\infty}
{n(L'_X)~ V_{max}(L'_X)~ \Psi(L'_X,L_X)}~ dL'_X
\end{equation}

where $n(L'_X)$ is the Schechter luminosity function and
$\Psi(L'_X,L_X)$ represents the Gaussian error distribution for 
the mean measurement uncertainty of 20\%. 

\begin{figure}
   \includegraphics[width=\columnwidth]{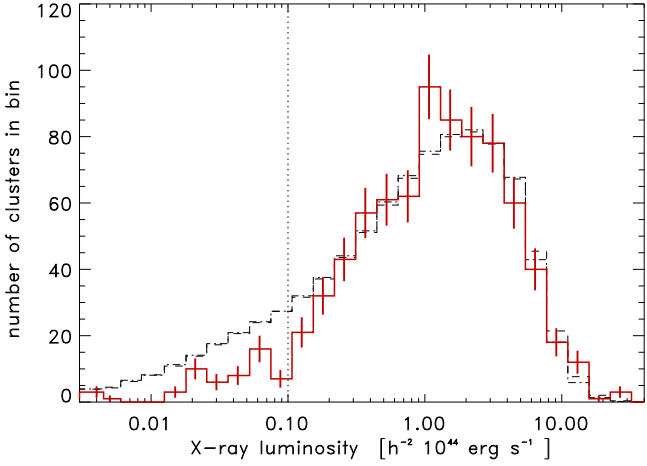}
\caption{X-ray luminosity histogram of the {\sf REFLEX II} clusters
at $z \le 0.3$  (red line with error bars) compared to the best fitting
Schechter function for the X-ray luminosity range $L_X \ge 0.1 \times
10^{44}$ erg s$^{-1}$. The X-ray luminosity limit is indicated by the 
vertical dotted line.
The error bars give the Poisson uncertainty of the counts in the
bins.
}\label{fig4}
\end{figure}

\begin{figure}
   \includegraphics[width=\columnwidth]{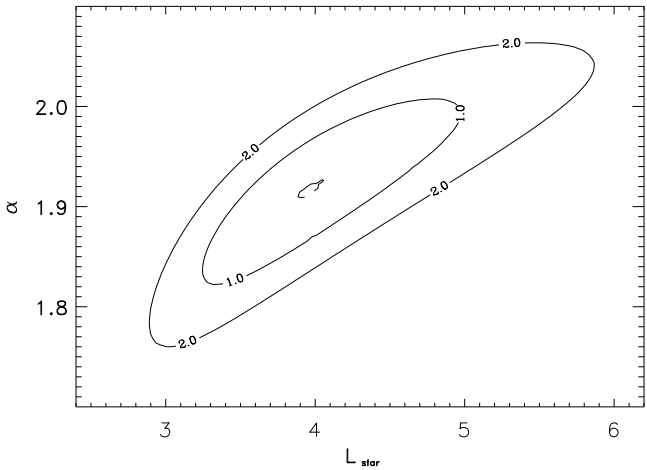}
\caption{Constraints on the Schechter parameters $\alpha$ and
$L_X^{\ast}$ from the fit to the data of the 
X-ray luminosity function of {\sf REFLEX II} for $z \le 0.3$
and  luminosity range $L_X \ge 0.1 \times 10^{44}$ erg s$^{-1}$.
}\label{fig5}
\end{figure}

Fig.~\ref{fig4} shows the luminosity distribution function and 
the fit 
for the redshift range $z \le 0.3$ and the luminosity
range $L_X  \ge 10^{43}$ erg s$^{-1}$.
The uncertainties of the fit for the two most important parameters,
$ \alpha$ and $L_X^{\ast}$ are shown in Fig.~\ref{fig5}, where we give
the 1 and 2$\sigma$ constraints. The typical $1\sigma$ uncertainties
-- similar for fits in different X-ray luminosity ranges -- are 
$\Delta \alpha = 5\%$ and $\Delta L_X^{\ast} = 22\%$.

The X-ray luminosity function 
is now observationally so well constrained, that a Schechter fit
is not any more an adequate description. Studying the fits for varying
X-ray luminosity ranges as shown in Table 1, 
we note that fits limited to the higher luminosity 
side of the luminosity function require a steeper slope, and the
fit is forced to a shallower slope for the full range of X-ray
luminosities, at the expense of a worse fit in the high and
intermediate luminosity range.

   \begin{table}
      \caption{Best fitting parameters for a Schechter function and
       alternative functions to the {\sf REFLEX II} X-ray luminosity function.}
         \label{Tempx}
      \[
         \begin{array}{llllllll}
            \hline
            \noalign{\smallskip}
 {\rm method} & ~~L_x-{\rm range}& ~~\alpha  & ~~L_X^{\ast} & ~~n_0 & q 
       & ~\eta & ~{\rm no. cl.} \\
            \noalign{\smallskip}
            \hline
            \noalign{\smallskip}
A & \ge 0.003  & 1.74  & 3.02 & 5.0\cdot 10^{-7} & - & 2.3 &  802 \\
A & \ge 0.01   & 1.76  & 3.10 & 4.7\cdot 10^{-7} & - & 2.5 &  798 \\
A & \ge 0.03   & 1.78  & 3.19 & 4.5\cdot 10^{-7} & - & 2.9 &  779 \\
A & \ge 0.1    & 1.93  & 3.99 & 2.8\cdot 10^{-7} & - & 9.9 &  741 \\
A & \ge 0.3    & 2.03  & 4.67 & 2.0\cdot 10^{-7} & - & 24.3&  623 \\
            \noalign{\smallskip}
            \hline
            \noalign{\smallskip}
B & \ge 0.003  & 1.54  & 1.08 & 1.8\cdot 10^{-6} & 1.39 & 1.4 &  802 \\
B & \ge 0.01   & 1.53  & 1.03 & 2.0\cdot 10^{-6} & 1.39 & 1.4 &  802 \\
B & \ge 0.1    & 1.60  & 1.17 & 1.6\cdot 10^{-6} & 1.39 & 2.5 &  802 \\
            \noalign{\smallskip}
            \hline
            \noalign{\smallskip}
C & \ge 0.003  & 2.19  & 5.7  & 1.3\cdot 10^{-7} & - & 1.6 &  802 \\
C & \ge 0.01   & 2.18  & 5.5  & 1.4\cdot 10^{-7} & - & 1.6 &  798 \\
C & \ge 0.03   & 2.13  & 5.0  & 1.7\cdot 10^{-7} & - & 1.4 &  779 \\
C & \ge 0.1    & 2.18  & 5.4  & 1.4\cdot 10^{-7} & - & 1.7 &  741 \\
C & \ge 0.3    & 2.17  & 5.4  & 1.4\cdot 10^{-7} & - & 1.6 &  623 \\
            \hline
            \noalign{\smallskip}
         \end{array}
      \]
{\bf Notes:} The method code $A$ is for the Schechter function (Eq. 2),
$B$ for the q-exponential (Eq. 5) and $C$ for the modified Schechter 
function (Eq. 6). The X-ray luminosity range used in the fit 
and the parameter $L_X^{\ast}$ are given in
units of $10^{44} h_{70}^{-2}$ erg s$^{-2}$. $\alpha$, $L_X^{\ast}$, and
$~n_0$ (in units of h$_{70}^3$ Mpc$^{-3}$)
are the slope, the break parameter and the normalisation of
the Schechter function or its modified variants, while $q$ is the 
extra parameter of the q-exponential function. $\eta$ is 
similar to a reduced 
$\chi^2$ parameter determined as explained in the text and the
last column gives the number of clusters involved in the fit.
The typical uncertainties are $\sim 5\%$ for $\alpha$ and 
$\sim 22\%$ for $L_x$. 
\label{tab1}
   \end{table}
%

\subsection{Alternative fitting functions}

Since the Schechter function does not provide such a satisfying 
analytical description of the {\sf REFLEX II} X-ray luminosity
function, we have also tried alternative functions for the fit.
As already used in our previous paper by Balaguera-Antolinez (2012)
we apply a q-exponential function of the following form:

\begin{equation}
{n(L_X)~dL_X}~ =~ n_0~ \left( {L_X \over L_X^{\ast}}\right)^{-\alpha}
\left[1~ +~ {L_X \over L_X^{\ast}}~ (1-q)\right]^{1 \over 1-q} {dL_X \over L_X^{\ast}}
\end{equation}

The effect of the q-exponential function is to cause a sharper decrease
of the Schechter function at high luminosities. This extra degree of
freedom helps to improve the fit as can be seen in Table 1.

To estimate how good the fit reproduces the data we calculate
the squared difference of the observed 
number of clusters in luminosity bins compared to the predicted 
using Eqs. 2, 5, and 6 nomalized by the Poisson error of the observations.
We derive a parameter denoted $\eta$ which is the
sum of the squared normalized differences over all bins,
devided by the number of bins. This is very similar to a reduced
$\chi^2$, with the difference that we do not normalize by
the degrees of freedom. We prefer to use the parameter $\eta$
since we want to characterize the deviation from the data rather
than the statistical significance of the fit.  
The summation is carried out
only over those bins with either a detection of at least one cluster
or a predicted number larger than 0.1. 
The values are given in Table 1. While the fit
is performed using data in the given luminosity range, the  $\eta$
parameter is always given for the full luminosity range.
One clearly notes the improvement of the fit for the q-exponential
function. Comparing the fit for the q-exponential function with
the earlier fitting results of Balaguera-Antolinez (2011), which
are converted to our cosmology model as $\alpha = 1.54$, $L_X^{\ast} = 1.2
\times 10^{44}~ h_{70}^{-2}$ erg s$^{-1}$, $n_0 = 1.4 \times 10^{-6}~ h_{70}^3$
Mpc$^{-3}$, and $q = 1.3$, we find a good agreement given the fact that
the sample was completed with additional cluster redshifts and the scaling relations
used were slightly improved.  

Probably an even better way to
arrive at a stable fit is to decrease the slope of the fitting function 
towards low luminosities. For this reason we also tried another modification 
of the Schechter function which bends the function down at very low 
luminosities. This modified function has the form:

\begin{equation}
{n(L_X)~dL_X}~ =~ n_0~ \left( {L_X \over L_X^{\ast}}\right)^{-\alpha}
exp\left(- {L_X \over L_X^{\ast}}\right)~ 
\left[1 -~\left(1+{L_x\over\beta}\right)^{-\gamma} \right] {dL_X \over L_X^{\ast}}
\end{equation}

where $\beta = 0.25$ and $\gamma = 1.7$ are extra parameters. Rather than
determining these two extra parameters in an overall fit, we found
suitable parameters before the final fit. This keeps the number of
fit parameters small, avoids degeneracies, and thus allows for a
better comparison of the different fitting results.

\begin{figure}
   \includegraphics[width=\columnwidth]{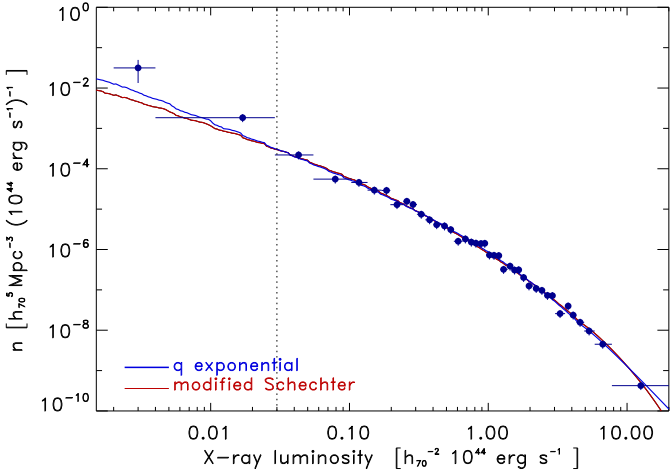}
\caption{X-ray luminosity function of {\sf REFLEX II} for a redshift
limit of $z = 0.3$ fitted by a q-exponential function and by the
modified Schechter function. Both fits were performed for the 
X-ray luminosity range $L_X \ge 0.03 \times
10^{44}$ erg s$^{-1}$ indicated by the vertical dotted line. 
}\label{fig6}
\end{figure}

We note that the fits are more stable and we basically recover
a good fitting function for the whole luminosity range even if
the fit is restricted to $L_X \ge 0.3 \times 10^{44}$ erg s$^{-1}$.
Therefore we consider the fitting results of this function as our
best description, at least for the luminosity range 
$L_X \ge 0.02 \times 10^{44}$ erg s$^{-1}$. The fits for the q-exponential
and the modified Schechter function are shown in Fig.~\ref{fig6}.
The fits do not very well describe the first two data points which
lay outside the fitted range, but note that these bins only contain 
23 clusters together and we put most emphasis on a good fit
for the rest of the data.  

\subsection{Comparison to previous determinations of the
X-ray luminosity function}
 
Previously determined XLF at low redshift generally show good 
agreement with the {\sf REFLEX I} XLF 
within the error limits of these smaller cluster samples.
This includes in particular the low redshift part of the RDCS, 
400 deg$^2$, and WARPS surveys (Rosati et al. 2002, Mullis et al. 2004, 
and Koens et al. 2013) and further references therein.
The {\sf REFLEX I}
survey is in good agreement with the present results. The Schechter
function fit presented for {\sf REFLEX I} in B\"ohringer et al. (2002)
is close to the present fit in the luminosity range $L_X \ge
3 \times 10^{42}$ erg s$^{-1}$. The small difference is that 
the {\sf REFLEX I} function
is slightly lower at small values of $L_x$, which is due to
a better sampling of the poor groups in {\sf REFLEX II}, and 
{\sf REFLEX I} slightly overshoots at high values of $L_X$ 
which is mostly due to a small decrease in the scaling radius for massive
clusters with the new scaling relations. 
These small changes do not affect the agreement with
previous results for the XLF.

\section{Model predictions of the cluster X-ray luminosity function}

The XLF of galaxy clusters can be used to obtain constraints
on cosmological model parameters, notably on the matter density parameter $\Omega_m$
and the amplitude parameter of the dark matter density fluctuations, $\sigma_8$.
For the determination of these parameters we need to compare the prediction of
the XLF for specific cosmological models with our observations.
In this section we describe our method for this prediction.

In the first step of the calculations we determine the cluster mass function based
on the recipe given by Tinker et al. (2008). A prerequisite for this calculation is the
specification of the statistics of the large-scale structure in the form of the dark
matter density fluctuation power spectrum. For the shape of the power spectrum
we assume that the initial power spectrum in the early Universe is described by a
power law with slope of 0.96 (consistent with the latest result from the {\sf Planck}
Mission (Planck Collaboration 2013b). We model the structure 
evolution to the present epoch by a transfer function as given
by Eisenstein \& Hu (1998)
including baryonic acoustic oscillations for a baryon density of $\Omega_b = 0.045$.
The amplitude of the power spectrum is specified by the amplitude parameter, $\sigma_8$.
This parameter is the variance of the fluctuation field filtered with a top-hat filter,
as shown in Eq.~7 below, with a radius of $8 h^{-1}$ Mpc. For the formula of the
mass function we calculate the variance of the filtered field through

\begin{equation}
\sigma^2(R_F) = {1 \over 2\pi^2} \int {P(k)~ \tilde{W}^2_{TH}(R_F,k)~~ k^2 dk}
\end{equation}

where $P(k)$ is the power spectrum at the epoch of consideration, 
$\tilde{W}_{TH}(R_F,k)$ is the top hat filter in Fourier space with 
filter radius $R_F$ in real space, 
and $\sigma(R_F)^2$ is the variance of the density fluctuation field. 
The variance as a function of filter radius can be transformed into 
a function of mass, using the mean density of the Universe, $\bar{\rho}_m$. 
The filter mass is given by $M = {4 \pi \over 3} \bar{\rho}_m R_F^3$ and
the mass function can then be written by

\begin{equation}
{ dn \over dM}~ =~ f\left[\sigma(M)\right]~ {\bar{\rho} _m  \over M}~ {d ln \sigma ^{-1} 
 \over dM}   
\end{equation}

with

\begin{equation}
f\left[\sigma(M)\right]~ =~ A \left[ \left({\sigma \over b} \right)^{-a} + 1 \right]~
exp  \left({c \over \sigma^ 2}\right)  ~~~~~.
\end{equation}

Here $M$ is the mass of the dark matter halos or clusters in terms of an
overdensity of 180 over the mean density of the Universe, $\bar{\rho} _m$.
We use for the overdensity the value of 180 and the following values for
the open parameters in Eq. 9, $A = 0.186$, $a = 1.47$, $b = 2.57$,
and $c = 1.19$ as given in Table 2 of Tinker et al. (2008).  
Some of these parameters are assumed to have a redshift dependence.
We use the following parametrizations, $A(z) = A_0 (1+z)^{-0.14}$,
$a(z) = a_0 (1+z)^{-0.06}$, and 
$b(z) = b_0 (1+z)^{-\alpha }$ with $log \alpha =  \left[{0.75
\over log( \Delta / 75)} \right] ^{1.2}$.

The mass given by this equation is the mass inside a mean overdensity of 
180 above the mean density of the Universe. We will be using masses 
defined for a mean overdensity of 200 above the critical density of the 
Universe at the epoch of light emission of the observed object.
Therefore we have to transform the mass equation into our definition.
For the conversion we assume that the mass profile of all clusters
can be described by a NFW-model profile (Navarro, Frenk \& White 1995,
1997) with a concentration parameter of 5.

Note that we will be using the following conventions throughout the 
paper. We will use an overdensity of 200 over the critical density of the
Universe to characterize the mass of the clusters (which is actually
an intermediate parameter not essential for the final results), because
this mass is closer to what we usually understand as the virial mass.
The X-ray parameters, like the X-ray luminosity are given inside a radius
of $r_{500}$, because this is closer to the observed aperture in which
X-ray luminosity is observationally determined. We do not introduce
an inconsistency by using different fiducial radii for mass
and luminosity as long as we have a careful 
book-keeping of all the convertions and scaling relations.

We then use an empirical mass - X-ray luminosity scaling relation 
of the form

\begin{equation}
L_{500}{\rm (0.1 - 2.4 ~keV)} = 0.1175~ M_{200}^{\alpha_{sl}}~ 
h^{\alpha_{sl}-2} ~~~E(z)^{\alpha_{sl}}
\end{equation} 

where $\alpha_{sl}$ is the slope of the scaling relation and $E(z) = H(z)/H_0$ is the 
evolution parameter related to the Hubble constant. 
The unit for $L_{500}$ is $10^{44}~ h_{70}^{-2}$ erg s$^{-1}$ (0.1 - 2.4 keV) and for
$M_{200}$ it is $10^{14}~ h_{70}^{-1}$ M$_{\odot}$. 
The redshift evolution of this relation is based on the assumption
of no evolution of the $L_X - T$ relation. For a discussion of this choice see
B\"ohringer et al. (2012). 
We also take into account that this relation
has an intrinsic scatter with a preferred value of $log\sigma_{L_X} = 0.114$ 
which corresponds to about $30\%$. We further fold in an observational error of
the luminosity of $20\%$ which is the fractional mean error in the
flux determination of the {\sf REFLEX II} cluster sample (B\"ohringer
et al. 2013).

For the slope of the $L_X - M$ scaling relation, $\alpha_{sl}$, we explore two
values, 1.51 and 1.61 to illustrate our uncertain knowledge. The value
of 1.61 is motivated by the results of our {\sf REXCESS} study (Pratt et al. 2009)
and the analysis of Vikhlinin et al. (2009). For {\sf REXCESS} a lower value of
1.53 is obtained before the assumed Malmquist bias correction. In comparison
to several other studies these values for the slope are on the high side
(see our survey of the topic in B\"ohringer et al. 2012). For example the complete
HIFLUGCS sample of 63 clusters gives a value of 1.46, the extended 
HIFLUGCS a value of 1.61
(Reiprich \& B\"ohringer 2002). Maughan (2007) finds 1.45 in
their cosmological analysis. Mantz et al. (2008) find a best fitting value of 1.24.
Therefore we prefer the value 1.51 as the best compromise for all the data
and for the relatively large and homogeneous HIFLUGCS sample as our prime
value.  

For the work in this paper we calculate the predicted luminosity function
for the median redshift of the {\sf REFLEX II} cluster sample of
$z = 0.102$ and quote the corresponding cosmological parameters for $z = 0$.
For the modeling of these two different epochs we use the standard
formulae for the linear growth of the density fluctuation field, since
it is the linear fluctuation evolution that is accounted for in
the applied statistical theory.

With these ingredients we can predict the X-ray luminosity function
analytically for different cosmological models. Note that for any 
variation of the cosmological model away from our reference model,
we have to convert the input X-ray luminosities, and the values of
$V_{max}$ with the corresponding survey selection function, and the
scaling relations to the new cosmology. 

\section{Cosmological parameters from the REFLEX X-ray 
luminosity function}

To determine the cosmological model parameters which best fit 
our observations we use a Likelihood function method in an analogous 
way to that used for the fitting of the Schechter function in
section 3.1. We are comparing again the predicted and observed
distribution function of X-ray luminosities, since this is very
close to an observable, as the spectral model 
assumptions used for their derivation introduce only negligible 
new uncertainties except for changes in the assumed geometry of 
the universe which are taken into account.
Thus we derive the following predicted distribution function
which now also accounts for the scatter in the
$L_X - M$ scaling relation 

\begin{equation}
\lambda (L_x)=
\int{n(L''_X)~ V_{surv}(L''_X)~ 
\Phi(L''_X,L'_X)~~ \Psi(L'_X,L_X)}~ dL''_X dL'_X 
\end{equation}

where $n(L_X) = {dn(L_X) \over dL_X} $ is the differential
X-ray luminosity function and $V_{surv}$
is the volume in which clusters with luminosity $L_X$ can be found.
In case of no redshift constraints the latter is equal to $V_{max}$, while
$\Phi(L''_X,L'_X)$ is the scatter in the $L_x -M$ relation and 
$\Psi(L'_X,L_X)$ is the uncertainty of the flux or luminosity
measurement.

\begin{figure}
   \includegraphics[width=\columnwidth]{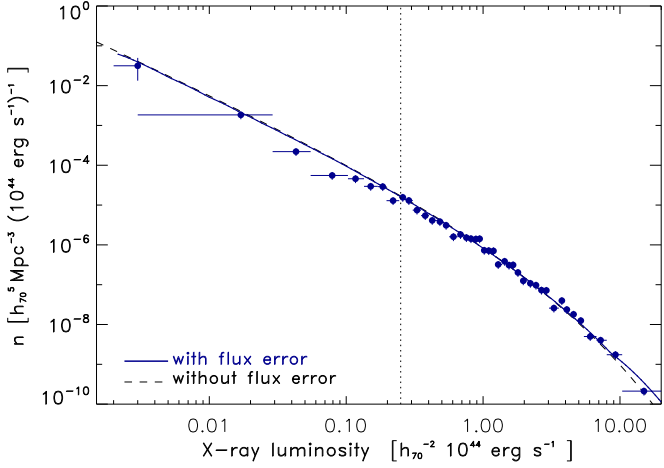}
\caption{
X-ray luminosity function of {\sf REFLEX II} (data points)
and the predicted X-ray luminosity function for the 
best fitting cosmological model. 
}\label{fig7}
\end{figure}

Fig.~\ref{fig7} shows the observed X-ray luminosity function
compared to the best fitting model prediction, 
where only clusters with $L_x \ge 0.25\times 
10^{44} h_{70}^{-2}$ erg s$^{-1}$ and $z \le 0.4$ have been 
used in the fit. This involves in total 698 clusters including the
lower total count limit of 20 source photons.
Over the fitted range the two functions show a very good agreement.
The reason for using a restricted luminosity range for this fit
is the fact, that for lower values of $L_X$, which corresponds to the
regime of galaxy groups, we have no reliable constraints on the
$L_X - M$ relation. It is too dangerous to rely on the assumption
that the relation is a straight power law over the full range of
observed X-ray luminosities.   

   \begin{table}
      \caption{Default cosmological model parameters.}
         \label{Tempx}
      \[
         \begin{array}{llr}
            \hline
            \noalign{\smallskip}
 {\rm parameter} & ~~~~~{\rm explanation}& {\rm ~~~~value}  \\
            \noalign{\smallskip}
            \hline
            \noalign{\smallskip}
h_{100} & {\rm Hubble~ parameter}  & 0.7 \\
\Omega_b & {\rm baryon~ density} &  0.045 \\
n_s & {\rm Primodial~ P(k)~ slope} & 0.96 \\
\alpha_{sl} & L_X - M~ {\rm relation~ slope}& 1.51 \\
n_0 & L_X - M~ {\rm relation~ norm.}  & 0.1175  \\
{\rm mass~bias} & {\rm X-ray~mass~underestimation} & 0.1 \\
            \hline
            \noalign{\smallskip}
         \end{array}
      \]
{\bf Notes:} In addition to the parameters we assume the model
to describe a flat $\Lambda$CDM Universe.
\label{tab1}
   \end{table}
%

In this section we are primarily interested in studying the constraints
on those cosmological parameters, for which we can get the most
interesting information, that is $\Omega_m$ and $\sigma_8$. We
therefore keep other parameters fixed, but we use a Hubble constant of $h=0.7$,
a flat universe, a baryon fraction compared to the critical density
of 0.045, and a spectral index of the primordial matter density fluctuation
power spectrum of 0.96. In addition we assume the Universe to be flat,
described by a $\Lambda$CDM model, that is a Dark Energy universe with
an equation of state parameter $w = -1$. These values are chosen to
be consistent with the 9-year WMAP results and with the {\sf PLANCK} results
(Hinshaw et al. 2012, Planck Collaboration 2013b), except for the
Hubble constant which is reported to be determined as $67.8 \pm
0.77$ for the combined results in the {\sf PLANCK} publication.
We also use a mass bias factor of 0.9
which accounts for the effect that the X-ray determined mass, which
is the base of the $L_x - M$ relation in Eq. 10, may be biased low
compared to the true mass by about 10\%. This has been found in 
simulations (Nagai et al. 2007, Valdarnini \& Piffaretti 2010, 
Meneghetti et al. 2010) and 
in comparison to weak lensing results (Mahdavi et al. 2013, Zhang et al. 2010,
Okabe et al. 2010). The assumed parameters are summarized in Table 2.
We explore the effect of some of these assumptions in a subsequent section.

\begin{figure}
   \includegraphics[width=\columnwidth]{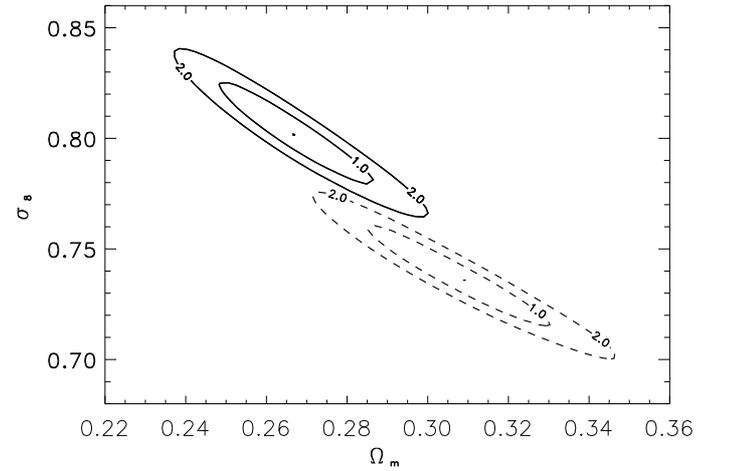}
\caption{Constraints of the cosmological model parameters $\Omega_m$
and $\sigma_8$ from the comparison of the observed {\sf REFLEX II} 
X-ray luminosity distribution and the cosmological model predictions.
Two results are shown for an $L_X-M$ relation slope of
1.51 (upper ellipses) and 1.61 (lower dashed ellipses). The contour lines show 
1 and $2\sigma$ confidence limits.
}\label{fig8}
\end{figure}

With these model assumptions the best constraints for the 
parameters $\Omega_m$ and $\sigma_8$ are shown in Fig.~\ref{fig8}.
We show two results for the significance limits, one for the preferred slope 
of the $L_x - M$ scaling relation of 1.51 and a second set of contours
for a slope value of 1.61.  
Table 3 summarizes the fitting results.
The uncertainties quoted here are only the statistical uncertainties 
for the constraints on these two parameters alone.
There is some degeneracy of the two parameters indicated by
the elongated error ellipses, but due to the well described
shape of the X-ray luminosity functions both parameters can be
well constrained. Fig.~\ref{figA.2} in the Appendix provides
an explanation for the behavoir of the constraints of 
$\Omega_m$ and $\sigma_8$. An increase of $\Omega_m$ increases the
amplitude of the X-ray luminosity function over the entire
luminosity range, this can be partly compensated by lowering
$\sigma_8$, but this affects mostly the high luminosity part.
This partial compensation leads to the elongated error ellipses,
but also allows the degeneracy to be broken. 

   \begin{table}
      \caption{Fit parameters of the best fitting cosmological model 
       prediction to the {\sf REFLEX II} X-ray luminosity function.}
         \label{Tempx}
      \[
         \begin{array}{lllll}
            \hline
            \noalign{\smallskip}
 {\rm fit}& L_X-M {\rm ~slope} & \Omega_m & \sigma_8  \\
            \noalign{\smallskip}
            \hline
            \noalign{\smallskip}
 L_X \ge 0.25   &  ~~ 1.51 & 0.27 \pm 10\%  & 0.80 \pm 3.5\%\\
 L_X \ge 0.25   &  ~~ 1.61 & 0.31 \pm 10\%  & 0.73 \pm 3.5\% \\
 L_X \ge 0.25   & {\rm marginalized} & 0.29 \pm 14\%  & 0.77 \pm 9\% \\
            \noalign{\smallskip}
            \hline
         \end{array}
      \]
{\bf Notes:} The first column shows the luminosity range used 
in the fit, while the redshift range is $z \le 0.4$. The second
column gives the slope of the $L_X - M$ relation used in the 
fit for the model prediction of the XLF.
For the results in the third line we have marginalized over the 
slope and normalisation of the scaling relation. 
\label{tab1}
   \end{table}
%

The lower panel of Fig.~\ref{figA.2} in the Appendix shows the 
changes of the predicted XLF with the change of the slope
of the $L_x - M$ relation. A steeper slope of the relation
results in a shallower slope of the X-ray luminosity function.
From this behaviour we can predict the change in the best fitting
cosmological parameters. A steeper slope of the $L_x - M$ relation
can be counteracted by lowering $\sigma_8$ with an additional higher 
value of $\Omega_m$ to adjust the normilazation. This is what we observe
in Fig.~\ref{fig8}.

\section{The role of scaling relations}

\begin{figure}
   \includegraphics[width=\columnwidth]{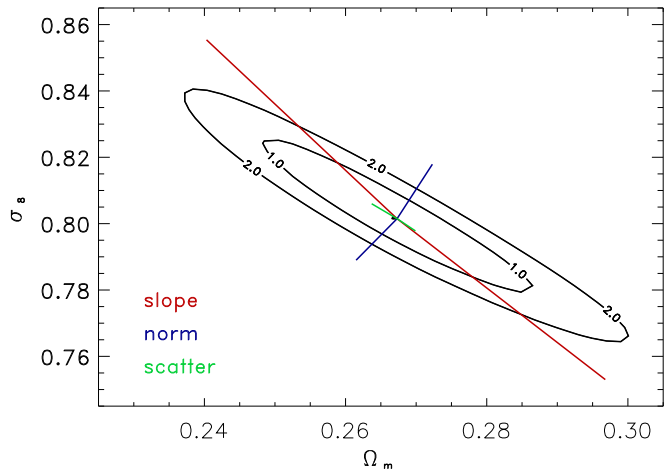}
\caption{Variation of the best fitting parameters for $\Omega_m$
and $\sigma_8$ with a change in slope, scatter and normalization
of the $L_X - M$ scaling relation. The end and mid points of the bars 
give the results for the following values, respectively:
slope = 1.435, 1.51, 1.586 ($\pm 5\%$, largest bars), normalization
= 0.1292, 0.1175, 0.1058 ($\pm 10\%$, bars in direction of minor axis),
scatter = 27\%, 30\%, 33\% ($\pm 10\%$, smallest bars). 
}\label{fig9}
\end{figure}

In the last section we already discussed the influence of the 
slope of the scaling relation on the cosmological fitting results.
A closer inspection of the effect of varying the different model
parameters for the fits to the observational data clearly
shows that the uncertainties in the observable - mass scaling relation 
is the most severe bottle-neck in deriving cosmological constraints
from our cluster survey. In Fig.~\ref{fig9} and Table 4 we show
the effect of the different parameters characterizing the  
$L_X - M$ scaling relation used in our modelling. The parameters
are the normalization, the slope, and the scatter in the relation.
A change in the slope has clearly the largest effect. It moves
the result in a direction very close to the major axis of
the error ellipse. A change in the normalization has a less
dramatic consequence, but it moves the result in an almost 
perpendicular direction, more along the minor axis of the 
error ellipse. Also the scatter has a non-negligible effect on the results.
We list the changes in the best fitting model parameters as
an effect of changes of the scaling relations in Table 4 in terms
of fractional changes and logarithmic derivatives. Since the effect
of the uncertainty in the measured flux is the same as that of
the scatter of the scaling relations (see Eq. 11), we also give 
the result for the fractional change of the combination of the 
two parameters (assuming that Gaussian error addition applies). 

To include these uncertainties for the slope and amplitude of the
scaling relation as the most critical parameters into our 
cosmological constraints, we perform a fit marginalizing over both 
parameters. We allow 
approximate $1\sigma$ uncertainties of 7\% for the slope and
14\% for the normalization. The uncertainty in the mass bias is
equivalent to a corresponding uncertainty in the normalization.
Thus this error includes both, the amplitude error of the relation
of $\sim 10\% $ and a 10\% mass bias uncertainty. The uncertainty
of $\pm 7\%$ for the slope of the scaling relation is motivated by
the difference of the most reliable results in the literature:
the statistically complete HIFLUGCS sample with $\alpha_{sl} = 1.46$,
Pratt et al. (2009) BCES (Y$\vert$X) fit with $\alpha_{sl} = 1.53$, and
Vikhlinin et al. (2009) with  $\alpha_{sl} = 1.61$. These studies
use galaxy cluster samples constructed to be representative of
the relevant survey population. We thus take the mean difference of 
these results as the sigma for the slope variation. We also
note that the statistical uncertainties of the slopes quoted in these
works is of similar magnitude as this difference.  
 
\begin{figure}
   \includegraphics[width=\columnwidth]{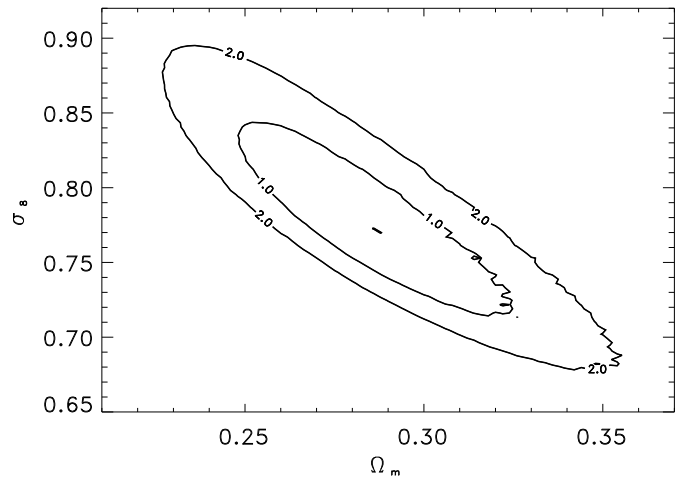}
\caption{Constraints of the cosmological model parameters $\Omega_m$
and $\sigma_8$ for the {\sf REFLEX II} cluster sample with a marginalization
over the most influential model parameters involving
the slope, $\alpha_{sl}$, and the normalization, $n_0$,
of the $L_X - M$ relation. Shown are the 1 and $2\sigma$ 
confidence limits. 
}\label{fig10}
\end{figure}

Fig.~\ref{fig10} shows the new constraints with relaxed assumptions
for the two most critical parameters. The contours enclose a larger 
area than those of Fig.~\ref{fig8}. The $2\sigma$ contour is now embedding
both error ellipses shown in Fig.~\ref{fig8}. One behaviour is noteworthy. 
The fit of the X-ray luminosity distribution favours a slightly higher
value for the slope than 1.51 for the best match of the observed and
predicted X-ray luminosity distribution. This shifts the center of the
uncertainty ellipses slightly towards the lower right corner of the 
plot, the direction of the result for a larger slope. The numerical 
values of the results are given in Table 3.

\section{Other systematic uncertainties}

As already mentioned it is not the aim of this paper to provide a
comprehensive marginalization over all uncertainties involved in
our analysis since we like to defer this task to a following publication
which uses a more appropriate approach. We rather like to make the effect
of the isolated variation of the important parameters transparent.
Therefore we studied the dependence of the results for  $\Omega_m$
and $\sigma_8$ on the variation of other cosmological parameters.
The results are summarized in Table 4. A parameter with a large
effect is the Hubble constant, $H_0$. For constant $\Omega_m$ a change in
the Hubble constant changes the mean density of the Universe and thus
it also changes the filter scale (in Eq. 7) for a cluster of given mass.
This explains only part of the effect of $H_0$, as we found that a 
change in  $H_0$ cannot be compensated by a corresponding change 
in $\Omega_m$. The effect of $H_0$ is more complex. While the baryon density 
in the Universe, a possible curvature, and the equation
of state parameter $w$ have a very minor influence,
the slope of the primordial power spectrum, $n_s$, is also an important
parameter to take into account. Finally, the role of  $f_{mass}$ was already
discussed, since its effect is the same as that of changing the 
$L_X - M$ normalisation, as can be seen from Eq. 10.

   \begin{table}
      \caption{Dependence of the best fitting values for the cosmological
       parameters $\Omega_m$ and $\sigma_8$ on the choice of other model
       parameters.}
         \label{Tempx}
      \[
         \begin{array}{lrrrrr}
            \hline
            \noalign{\smallskip}
 {\rm parameter}& ~~~~~~~{d ln \Omega_m \over d ln P}& 
 ~~~~~~~{d ln \sigma_8 \over d ln P} & ~~~~~~~\Delta P & 
~~~~~~~\Delta \Omega_m & ~~~~~~~\Delta \sigma_8   \\
           \noalign{\smallskip}
            \hline
            \noalign{\smallskip} 
\alpha_{sl} & 2.2   & -1.3  &  10\% & 22\%  & 13\% \\
n_0        & -0.22 & -0.18 &  10\% & -2.2\% & -1.8\% \\    
{\rm scatter}& 0.12& -0.05 &  10\% &  1.2\% & -0.5\% \\
{\rm scatter~+~ferr. } & 0.16 & -0.07 & 10\% &  1.6\% & -0.7\% \\ 
           \noalign{\smallskip}
            \hline
            \noalign{\smallskip}
 H_0       &  -0.7   & 0.4  & 5\%  & -3.5\% &  2\% \\
 \Omega_b  &   0.15  & -0.07  & 10\% & 1.5\% &   -0.7\% \\
 n_s       &  -0.9   &  0.5 & 5\%  & -4.5\% &  2.5\% \\
 \Omega_m + \Omega_{\Lambda}  & \le \pm 0.11^a)   & \le 0.07  & 5\%  & \le \pm 0.5\% & \le 0.35\% \\
 f_{mass}   &  - 0.3   & - 0.28  & 10\%  & -3\% &  -3\% \\
  w        &  \le 0.02 & \ge -0.02 &  10\% &  0.15\% & -0.19\% \\
            \noalign{\smallskip}
            \hline
         \end{array}
      \]
$^a)$ $\Omega_m$ decreases with a change of curvature in both directions.\\
{\bf Notes:} The parameters for the $L_X - M$ scaling relation are 
the slope $\alpha_{sl}$, the normalization, $n_0$, and the scatter. 
We also show the effect of the variation of the combined scatter and 
flux error. The cosmological parameters 
in the second part of the Table are $H_0$, the Hubble constant, $\Omega_b$
the baryon density, and $n_s$ the slope of the primordial power
spectrum.  $\Omega_m + \Omega_{\Lambda}$ gives with the degree of the deviation
from one the deviation from a flat universe, $f_{mass}$ provides the
mass bias, the correction from the X-ray determined hydrostatic mass to the
true mass of the clusters, and $w$ shows the effect of a constant equation
of state parameter for Dark Energy. $P$ in the formulas for the columns indicates
the parameter of the row. $\Delta \Omega_m$ and $\Delta \sigma_8$ give
the change of the resulting values for these two cosmological parameters
upon a parameter variation of $\Delta P$. \\
\label{tab1}
   \end{table}
%

We have also checked the effect of sample incompleteness on the
results of the cosmological constraints by arbitrarily excluding a
fraction of the observed clusters from the sample analysed.
The effect of a reduction of the sample by 10\% for example decreases
$\Omega_m$ by 3.5\% and $\sigma_8$ by increases by less than 0.2\%. Thus it is clear
that at this stage the sample completeness, which is very high for 
{\sf REFLEX II}, is of no concern for the results discussed here.

An inspection of Table 4 shows again that the most dramatic effect
on the results comes from the uncertainty of the slope of the
$L_X - M$ scaling relation, while the effect of the normalisation
is among the next most important parameters together with
$H_0$ and $n_s$. Therefore we are confident that the marginalization
over these two most important parameters gives a fair and conservative
account of the overall uncertainty of our cosmological constraints.

The relatively smaller effect of these other cosmological parameters
on the $\Omega_m$  and $\sigma_8$ parameters compared to the uncertainties of
the scaling relation was e.g. also discussed in Voevodkin \& Vikhlinin (2004),
Vikhlinin et al. (2009), Henry et al. (2009) and Rozo et al. (2010)
with the same conclusion.

\section{Discussion and comparison to previous results.}

A comparison of the results on the cosmological parameter constraints shown
in Fig.~\ref{fig8} and Fig.~\ref{fig10} clearly shows how much information 
is lost due to our imprecise knowledge of the observable - mass scaling relations.
Therefore one of our major efforts in the future will be invested to improve
this situation.

One effect not included in the error bars of the XLF given here is
the sample variance. This effect is caused by structure on very large
scales such that our survey results would be slightly different would the
data have been sampled in exactly the same way in a different region
of our Universe. We estimated the cosmic variance in the same way as
done in our previous paper of the XLF of {\sf REFLEX I} (B\"ohringer et al. 2002)
by approximating the survey volume by a sphere. We then calculate the
variance of the dark matter density from the power spectrum filtered
by a top hat filter with the size of this sphere. This approach and 
its effects are also described by Hu \& Kravtsov (2003). To further
estimate the variance effect on the cluster density we also have to 
account for the fact that the cluster density is biased with respect
to the dark matter. The bias factor depends on the X-ray luminosity
or the mass of the clusters. The calculations of the bias
are based in the formulas by Tinker et al. (2010)
and have been described in Balaguera-Antolinez et al. (2011).
We find that the sample variance dominates the uncertainties in the
cluster density in the XLF at $L_X < 8 \times 10^{42}~ h_{70}^{-2}$ erg
s$^{-1}$. It is less than half at
$L_X \sim 2.5 \times 10^{43}~ h_{70}^{-2}$ erg s$^{-1}$,  the lowest luminosity
included in our fits, and it decreases rapidly thereafter. Thus
we have not corrected for this effect in the present work.

With our conservative approach of marginalization over the most
important systematic uncertainties, we can now compare our results to 
other cosmological studies using galaxy clusters. The result most 
similar in the nature of the constraints is the recently published 
result obtained with galaxy clusters detected
in the {\sf PLANCK} microwave background survey through the Sunyeav-Zel'dovich
effect (SZE, Planck Collaboration 2013a). This survey is also an all-sky
survey. There are 189 galaxy clusters detected at high signal/noise used
for the cosmological analysis. 92\% of the clusters have redshifts
$z \le 0.4$ and 81\% have $z \le 0.3$ compared to our 
{\sf REFLEX II} sample where
99\% have $z \le 0.4$ and 92\% $z \le 0.3$. Thus even though the
{\sf REFLEX II} sample has on average lower redshifts, the redshift range
is still quite similar. The cosmological analysis for the
{\sf PLANCK} clusters uses a different approach than the one used here.
While we have reproduced the luminosity distribution, in the  
{\sf PLANCK} analysis the redshift distribution is used to determine 
the best fit. In the absence of strong evolutionary effects in the 
galaxy cluster population for the small redshift range, the approaches
have in effect some similarity. In the probed redshift range in the nearby Universe,
the {\sf PLANCK} cluster sample is approximately flux limited,
as the measured integrated SZE is proportional to the apparent
surface area of the objects which is decreasing with the inverse square of
the distance for low redshifts. A redshift histogram at low
redshift  thus reflects the cumulative luminosity function
probed at different luminosity cuts. In this sense the two approaches 
are quite comparable.

\begin{figure}
   \includegraphics[width=\columnwidth]{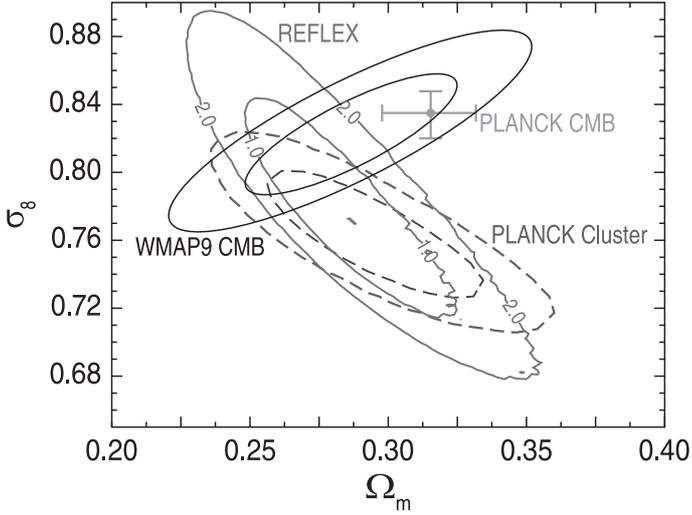}
\caption{Comparison of the marginalized cosmological model constraints 
for the {\sf REFLEX II} survey with the results from the {\sf PLANCK} Mission
for galaxy clusters (Planck Collaboration 2013a) as well as with the cosmic 
microwave background (CMB) power spectrum
(Planck Collaboration 2013b) and with the CMB results from the 9-year {\sf WMAP}
data (Hinshaw et al. 2013).  The contour lines give the 1 and 2$\sigma$ results
for the cluster surveys and {\sf WMAP} mission and the cross gives 
the  1$\sigma$ result from  {\sf PLANCK} CMB (Planck Collaboration 2013b).
}\label{fig11}
\end{figure}

There is another similarity between the two surveys. The mass-observable
relation for the SZE detected {\sf PLANCK} clusters relies heavily
on the calibration of the X-ray properties of the clusters (see
Arnaud et al. 2010, Melin et al. 2011, and Planck Collaboration 2013a). 
Therefore we actually expect to see constraint results which give
a very similar parameter range. Since systematic effects are more 
important at this current state, the larger size of the 
{\sf REFLEX II} sample does not play a major role. Therefore it 
is on one hand not very surprising to see such a good
agreement between
the cosmological constraints from the two surveys as shown in 
Fig.~\ref{fig11}. On the other hand the cluster detection and 
selection is very different and the agreement gives much support
to the different survey methods.

In Fig.~\ref{fig11} we also compare the cluster results for the constraints
on $\Omega_m$ and $\sigma_8$ to those derived from the analysis of
the microwave background anisotropies as measured with WMAP 9yr data (Hinshaw
et al. 2013) and by the {\sf PLANCK} mission (Planck Collaboration 2013b).
As discussed in detail in the {\sf PLANCK} result paper on cosmological constraints
from galaxy clusters (Planck Collaboration 2013a), the two results
from {\sf PLANCK}  show some discrepancy. Here we confirm the {\sf PLANCK} cluster results
from X-ray observations very well, using a somewhat different approach based
on the luminosity distribution instead of the redshift distribution of the 
clusters. In Fig.~\ref{fig11} we show the best cosmological fit
for {\sf PLANCK} CMB data with {\sf WMAP} polarization data
quoted in the {\sf PLANCK} cosmology paper (Planck Collaboration 2013b)  
with 1$\sigma$ error bars. Note that this result is closer to the cluster constraints 
than the error ellipses shown in Planck Collaboration (2013a).

\begin{figure}
   \includegraphics[width=\columnwidth]{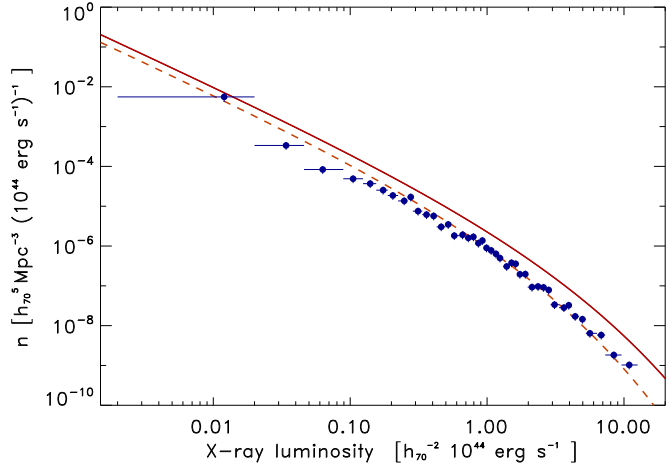}
\caption{Comparison of the observed 
{\sf REFLEX II} luminosity function (data points) with the theoretical prediction
for the best fitting model to the {\sf PLANCK} CMB data (Planck 
Collaboration XVI 2013b) with $\Omega_m = 0.315$ and
$\sigma_8 = 0.834$ with a mass bias of 10\% (upper solid line).
Same calculation but assuming a mass bias of 45\% is shown by the
lower dashed line.  
}\label{fig12}
\end{figure}

How difficult it is to make our results compatible with 
the {\sf PLANCK} CMB constraints is illustrated in Fig.~\ref{fig12}.
There we show the predicted X-ray luminosity function for the 
{\sf PLANCK} CMB cosmology compared to the 
one observed in the {\sf REFLEX} survey. The dramatic difference in cluster 
number counts is apparent. It was already discussed in the {\sf PLANCK}
paper on cosmological results from clusters (Planck Collaboration 2013a)
that the results could very roughly be reconciled by allowing a mass
bias of 45\% (that is assuming that the X-ray mass determination 
covers on average only 55\% of the true mass of a cluster).
This assumption is included in the result shown in 
Fig.~\ref{fig12}. It gives a rough agreement, but the value
for the mass bias is far outside the range of acceptable values.
In the comparison of X-ray determined masses with mass measurements 
from gravitational lensing (Mahdavi et al. 2013, Zhang 
et al. 2010, Okabe et al. 2010), one finds small biases with 
X-ray masses lower by few percent to about 20\%. The large mass bias needed
above is not compatible with these observational results.
We also note in Fig.~\ref{fig11} that the cluster result has an 
overlap with the WMAP CMB data.

\begin{figure}
   \includegraphics[width=\columnwidth]{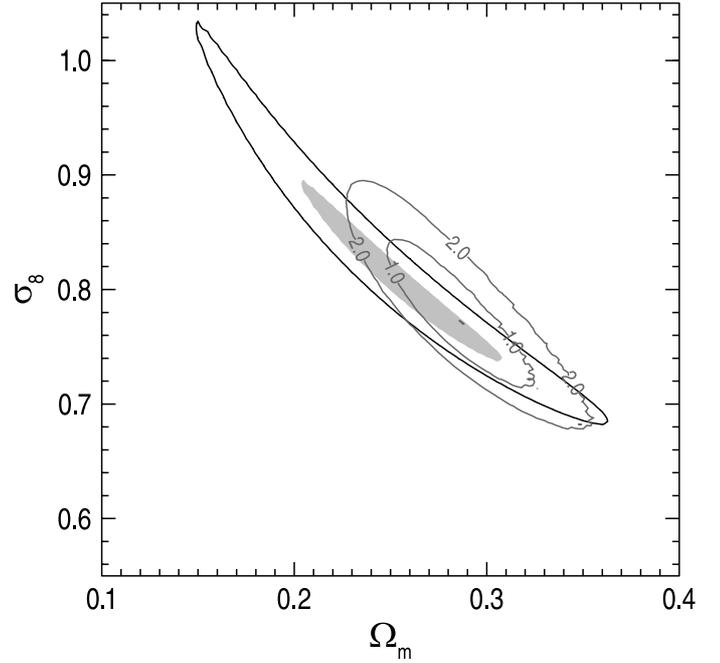}
\caption{Comparison to the results from the X-ray cluster survey
by Vikhlinin et al. (2009). The constraints on $\Omega_m$ and $\sigma_8$
shown here have been derived from the low redshift sample of 49
clusters at $<z> = 0.05$ of their survey. 1 and 2$\sigma$ contours
are shown for both surveys.
}\label{fig13}
\end{figure}

We can also compare our results with previous X-ray cluster studies
by Vikhlinin et al. (2009), Henry et al. (2009), and Mantz et al. (2008).
The cosmological analysis of Vikhlinin et al. (2009) is based on
a nearby sample of 49 clusters selected from the brightest clusters
in the sky with $<z> \sim 0.05$ and a more distant cluster sample
of 37 clusters from the 400 deg$^2$ survey (Burenin et al. 2007)
 with $<z> \sim 0.55$. All clusters are well observed with 
Chandra so that masses can be estimated
from low scatter mass proxies like gas mass, $M_{gas}$,
and the product of gas mass and bulk temperature, $Y_X$.
While the full set of data is used to obtain information on
the equation of state parameter of Dark Energy, $w$,
the parameters $\Omega_m$ and $\sigma_8$ are best constraint
by the low redshift sample only,
as discussed in their paper. In Fig.~\ref{fig13} we compare
their results to ours and find very good agreement. Our
results are better constrained along the degeneracy direction,
due to the better information we have on the shape of the
XLF.

Also the work by Henry et al. (2009) uses a small sample of local
clusters with a better mass proxy than X-ray luminosity, which is
the spectroscopically measured intracluster medium temperature
from X-ray observations. Their sample comprises 48 of
the brightest clusters in the sky (from HIFLUGCS, Reiprich \& B\"ohringer
2002) with temperatures measured with the ASCA Satellite taken
from Horner et al. (2001) and Ikebe et al. (2002) with $z \le 0.2$
and $k_BT \ge 3$ keV. The results 
obtained by marginalizing over 12 uninteresting parameters are
characterized by very elongated error ellipses (see their Fig. 10) for which the 
authors provide the parameterization, $\sigma_8~ \left({\Omega_m \over 0.32}
\right)^{0.3} = 0.86 (\pm 0.04)$. Our results shown in Fig.~\ref{fig10}
are best characterized by $ \sigma_8~ \left({\Omega_m \over 0.27}
\right)^{0.57} = 0.80 (\pm 0.03)$. Transforming both relations to the
same pivot point in the middle, we find: $\sigma_8~ \left({\Omega_m \over 0.30}
\right)^{0.3} = 0.877 (\pm 0.04)$ for the results of Herny et al. (2009)
and $ \sigma_8~ \left({\Omega_m \over 0.30} \right)^{0.57} = 0.753 (\pm 0.03)$
for our work. It displays an offset between the two results of almost 
$2 \sigma$. The data used by Herny et al. (2009) stem from the 
pre-Chandra/XMM-Newton era without good spatially resolved 
spectroscopy. Thus the small difference between these results and 
ours is not a reason for concern.

Mantz et al. (2008) use for their analysis 130 clusters from {\sf REFLEX I}
with a lower luminosity cut of $L_X = 2.55 \times 10^{44} h_{70}^{-2}$ erg s$^{-1}$
in the 0.1 - 2.4 keV band, 78 clusters from the northern BCS sample,
and 34 clusters from MACS with a flux limit of $F_X = 2 \times 10^{-12}$
erg cm$^{-2}$ s$^{-1}$. The cosmological constraints on the two parameters
$\Omega_m$ and $\sigma_8$ from an analysis of the cluster data alone
marginalizing over a comprehensive set of other parameters is shown in
their Fig.\ 7. Our error ellipse from  Fig.~\ref{fig10} falls nicely in 
the middle of their constraints region. At $\Omega_m = 0.27$ their constraints
are $\sigma_8 = 0.805 \pm 0.1$ for example in very close agreement to our 
results. 

\begin{figure}
   \includegraphics[width=\columnwidth]{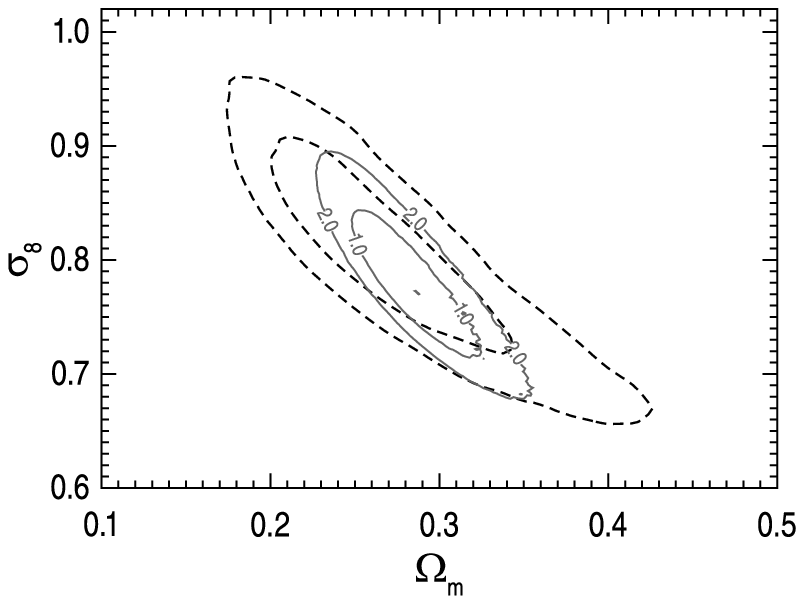}
\caption{Comparison to constraints from SDSS MaxBCG clusters 
(Rozo et al. 2010). The more than 10 000 clusters cover a redshift
range of $z = 0.1 - 0.3$ and an approximate mass range of 
$M = 7 \times 10^{13} - 2 \times 10^{15}$ M$_{\odot}$. 
}\label{fig14}
\end{figure}

Rozo et al. (2010) derived constraints for cosmological parameters from
an analysis of the galaxy cluster population in the Sloan Digital Sky Survey
(SDSS) obtained by means of the MaxBCG cluster detection method (Koester et al. 2007).
10810 galaxy cluster are used in the redshift range $z = 0.1 - 0.3$ from
a survey area of 7398 deg$^2$ of SDSS-DR4+. While the number of clusters of
the sample is much larger than that from any other survey discussed here, the scatter
in the richness - mass relation is much larger than for the X-ray related
mass proxies. The richness mass relation is calibrated by an analysis
of gravitational lensing data of stacked cluster shear profiles from SDSS.
In Fig.~\ref{fig14} we compare our results to their constraints
noting a very good overlap.

\begin{figure}
   \includegraphics[width=\columnwidth]{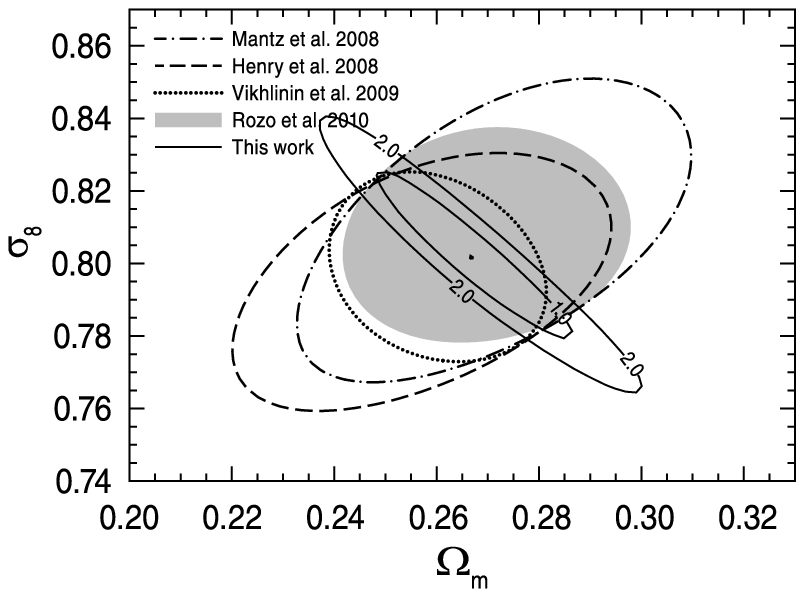}
\caption{Comparison to optical and X-ray surveys, where the data have been 
combined with WMAP 5yr data (Rozo et al. 2010, Mantz et al. 2008, Vikhlinin
et al. 2009, Henry et al. 2009). Our data are the {\sf REFLEX} constraints
with no marginalization over scaling relation systematics. The
contours for the literature data give 68\% confidence limits and
for the {\sf REFLEX II} data 1 and 2$\sigma$ contours are shown.
}\label{fig15}
\end{figure}

In Fig.~\ref{fig15} we then compare our results to
a set of the X-ray and optical results
discussed above, where the constraints from the literature 
have further been tightened by 
combining the cluster constraints with those of WMAP 5yr data (Dunkley
et al. 2007). For the display of our results we use our preferred 
$L_X - M$ scaling relation from Fig.~\ref{fig8} neglecting the 
uncertainties of this relation.
We note that this relation shows prefect consistency with the other 
cosmological data. Thus using the further constraints from the
cosmic microwave background anisotropies we are pointed towards an
observable - mass relation for our clusters which agrees very well with
the default relation used in this paper. 

In our work, the good characterization of the shape of the XLF
is the means to break the degeneracy between the  $\sigma_8$ and 
$\Omega_m$ parameter. Another way to get a better separation between
the two parameters is to extend the cluster study to higher redshifts. This
has been recently achieved in ground based microwave surveys, detecting
clusters in the SZE. Thus to compare our findings to their results
provides another independent consistency check for our approach.
From the cluster sample detected with the South Pole Telescope (SPT)
Benson et al. (2013) obtained cosmological parameter constraints from
18 massive clusters in the redshift range $ z=0.3$ to  $1.1$. Our constraints
fall very nicely into the middle of their constraints using 
an $H_0$ prior and Big Bang Nucleosynthesis results. Our results
also overlap well with the combined constraints of SPT cluster
and WMAP CMB data. The other distant SZE cluster sample comes from
the microwave background survey with the Atacama Cosmology Telescope
(ACT). Cosmological constraints have recently been derived by
Hasselfeld et al. (2013) from 15 massive clusters in the redshift range $z = 
0.2 - 1.4$. Our results overlap well with their constraints
in the lower half of their contours at lower $\sigma_8$.

\section{Summary and conclusions}

We used the {\sf REFLEX II} catalog and the well defined selection function
to construct the X-ray luminosity function for the sample at a median
redshift of $z = 0.102$. For the flux limit of $F_X = 1.8 \times 10^{-12}$
erg s$^{-1}$ cm$^{-2}$, a luminosity limit of $L_X = 0.03 \times 10^{44}$
erg s$^{-1}$ (for 0.1 - 2.5 keV), a lower photon count limit of 20 source photons,
and a redshift range of $z = 0 - 0.4$, some 819 clusters are included
in the luminosity function determination. The cluster catalog
is better than 90\% complete, with a best estimate of about
95\% as detailed in B\"ohringer et al. (2013) and we also expect a few percent
contamination by clusters whose X-ray luminosity may be boosted by an AGN.
Since this fraction is small and as an incompleteness even of the order of
10\% causes only little change in the derived cosmological parameters as detailed
in section 7, we have not included any correction for incompleteness in the
presented results.  

Inspecting the XLF in different redshift shells reveals no
significant evolution of the XLF. We showed for our best fitting cosmological
model that this undetectable change is consistent with the theoretical expectation.
This does not imply that there is no evolution in the cluster mass function.
There are two competing effects, the evolution of the mass function
and the evolution of the $L_X - M$ relation. The relation of
X-ray luminosity to mass evolves as clusters have been 
more compact on average in the past which increases the X-ray luminosity
(through the square-law dependence on the density) and clusters of the same mass 
become brighter. This compensates for the loss of massive clusters at higher redshifts
and suppresses the evolution in X-ray luminosity.

In search of a good analytical description of the XLF, we found that the
Schechter function does not capture our knowledge with sufficient precision.
We therefore propose a modified Schechter function for a good description
of the data.

The most interesting application of the XLF is to test theoretical predictions
of this function within the frame of different cosmological models. These tests
are based on the theory of cosmic evolution and structure formation and rely
among other things on the description of the transfer function of the power
spectrum by Eisenstein \& Hu (1998), on the numerical simulation calibrated
recipe for the cluster mass function by Tinker et al. (2008), and on scaling 
relations that enable the connection of cluster mass and X-ray luminosity.
Apart from the scaling relations, the other theoretical framework has been
intensively tested and is believed to be accurate at about the 5\% level.

We find that we can get a very good match of the observed XLF with the
theoretical predictions for a very reasonable
cosmological model; in particular if we restrict the fitting to the
luminosity range $L_X \ge 0.25 \times 10^{44}$ erg s$^{-1}$ (0.1 - 2.4 keV)
where we have an observationally calibrated $L_X - M$ relation (note the
good match of the predicted and observed XLF in Fig.~\ref{fig7}).
In using the observational data to constrain cosmological parameters
we have in this paper not pursued a comprehensive marginalization over all
relevant parameters. We postpone this to later work. We rather wanted
to gain an overview and an understanding of the effect of the different 
parameters by studying them individually. From this investigation it becomes
clear that by far the largest uncertainty in the constraints of cosmological
parameters is introduced by the $L_X - M$ scaling relation and specifically
its slope and normalization. We give a detailed account on the influence of the
other parameters and then concentrate on a marginalization study including these two
most important input parameters, which provides a good account of the overall
uncertainties. The important constraints that we derive from the {\sf REFLEX II} 
data are in the matter density parameter $\Omega_m = 0.29 \pm 0.04$
and the amplitude parameter of the matter density fluctuations 
$\sigma_8 = 0.77 \pm 0.07$.

The currently most interesting comparison of our findings with other results
is that with the recently published cosmological constraints from clusters 
detected with {\sf PLANCK} (Planck Collaboration 2013b). The {\sf PLANCK} results 
show a tension between the cosmological constraints on $\Omega_m$ 
and $\sigma_8$ from clusters and from the cosmic microwave background (CMB)
anisotropies, which has caused a lively debate. We find that our results
agree perfectly with the {\sf PLANCK} cluster data and it would be very hard
to reconcile them with the CMB derived results. However, we find that
our results are consistent with the constraints from the CMB study with WMAP
(Hinshaw et al. 2013). Since there is also some tension between the 
implications from the CMB data of WMAP and {\sf PLANCK}, the source of which is
currently under investigation, we are confident, that the solution of these
problems will bring a closer agreement of all the data in the future without
a significant change of our results.

The good agreement of our results with the recent work on cluster cosmology
in the literature is encouraging. But we should point out here that 
our new results provide tighter constraints on the two tested parameters
than the previous studies and constitute a significant improvement.
We have, however, reached a limit where a further increase of the clusters
sample and of the overall statistics will not lead to much further improvements,
if we cannot better calibrate the scaling relations. A major reason
for our poor knowledge on the scaling relations originates in several facts.
On one hand the cluster samples with very well defined selection criteria
used in the scaling relation studies are still very small with typically
30 - 50 objects. Another source of uncertainty is revealed by the difference
in results for mass, temperature, or X-ray luminosity determined for 
the same set of clusters by different authors (e.g. Reichert et al. 2011). 
And there are still some
calibration uncertainties for the XMM-Newton and Chandra instruments
for which there is an ongoing effort of their resolution (e.g. Kettula et al. 2013). 
Therefore one of the next major efforts of the authors will be 
to increase the sample size and the data reduction quality of the
cluster samples to obtain better constraints on the important
scaling relations.

\begin{acknowledgements}
We like to thank the {\sf ROSAT} team at MPE for the support with the data
and advice in the reduction of the RASS and the staff at ESO La Silla
for the technical support during the numerous observing runs conducted
for the {\sf REFLEX} project since 1992. Special thanks goes to Peter Schuecker,
who was a very essential part of our team and who died unexpectedly in 2006.
H.B. and G.C.  acknowledge support from the DFG Transregio Program TR33
and the Munich Excellence Cluster ''Structure and Evolution of the Universe''.  
C.A.C. acknowledges STFC for financial support.
\end{acknowledgements}

\appendix
\section{Expected evolution of the X-ray luminosity function}

\begin{figure}
   \includegraphics[width=\columnwidth]{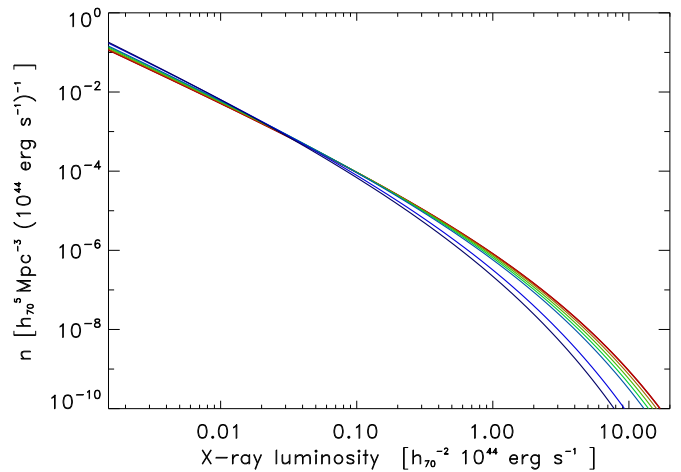}
 \caption{Predicted evolution of the X-ray luminosity function of 
galaxy clusters. The curves from top to bottom
show the XLF for $z=0$ , 0.1, 0.2, 0.3, 0.4 0.8, 1.0, respectively.
}\label{figA.1}
\end{figure}

Since we could not detect a significant evolutionary effect of the
luminosity function as a function of redshift in the {\sf REFLEX II} sample
in our analysis in this paper, we take a closer look at the expected 
evolution. The evolution is driven by the building up the mass of clusters
increasing the high mass end of the cluster mass function with time.
But this evolutionary effect is partly conteracted by the evolution
of the $L_X - M$ relation, so that the X-ray luminosity finction
shows very little evolution at low redshifts.This is illustrated in
Fig.~\ref{figA.1} which shows the predicted X-ray luminosity function
at redshifts from $z = 0$ to 0.4 and at $z = 0.8$ and 1. 
We note that unless we can probe the high mass
end of the X-ray luminosity function very precisely at low redshift,
we cannot not probe for this small effect. Due to the small volume of 
the survey at low redshifts we do not have the statistics to probe
the high mass end sufficiently and thus we have no good leverage
to constrain the luminosity function evolution. This does not imply,
however, that there is no evolution of the cluster mass function.

\section{Shape dependence of the XLF on various parameters}

\begin{figure}
   \includegraphics[width=\columnwidth]{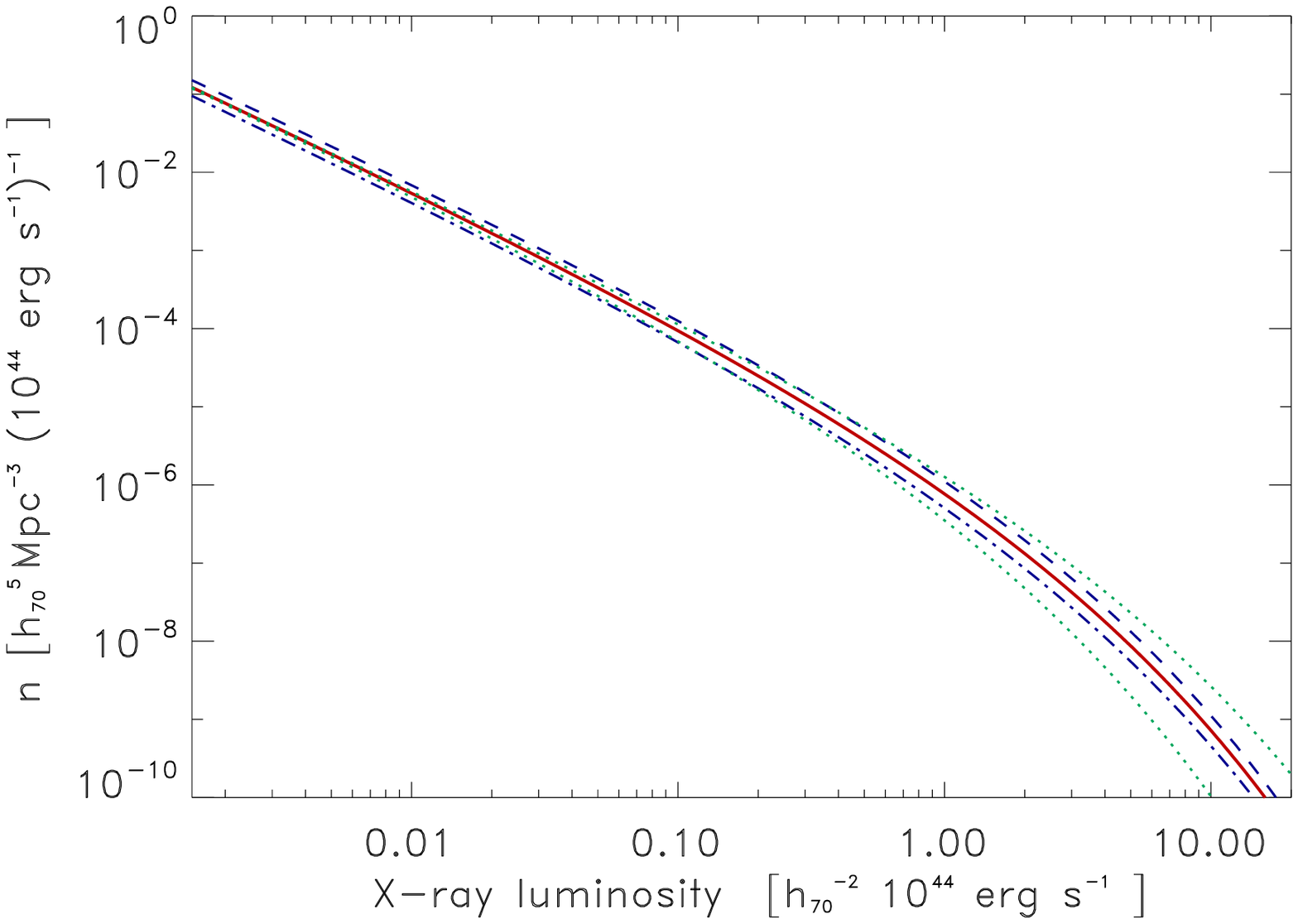}
   \includegraphics[width=\columnwidth]{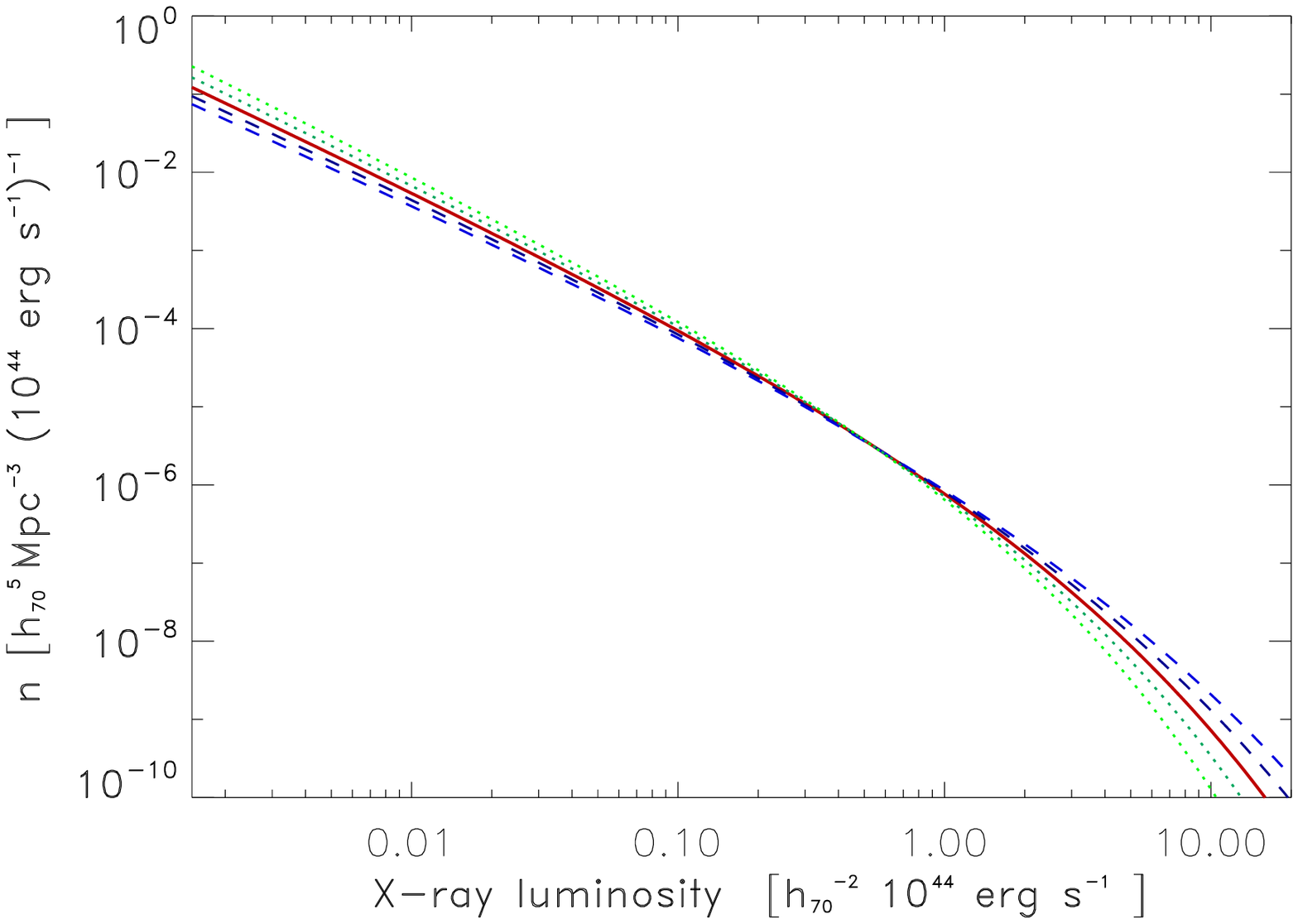}
\caption{{\bf Upper panel:} Variations of the predicted XLF with the change in the
parameters $\Omega_m$ and $\sigma_8$. The reference model (solid red
line) has $\Omega_m = 0.26$ and $\sigma_8 = 0.8$, the dashed line
has $\Omega_m = 0.30$ and the dot-dashed line $\Omega_m = 0.22$,
while the upper dotted line shows the result for $\sigma_8 = 0.9$
and the lower one for $\sigma_8 = 0.7$.
{\bf Lower panel:}
Variations of the predicted XLF with the change in the
parameter, $\alpha_{sl}$ of the $L_X - M$ scaling relation.
The reference model (solid red line) has $\alpha_{sl} = 1.51$, the two
dashed lines correspond to $\alpha_{sl} = 1.61$ and $1.71$ and the 
two dotted lines to $\alpha_{sl} = 1.41$ and $1.31$.
}\label{figA.2}
\end{figure}

In this section we show how the XLF depends on the cosmological parameters
$\Omega_m$ and $\sigma_8$ on the one hand and on the slope and
normalization parameter of the $L_X - M$ scaling relation on the
other hand to explain the
behaviour of the parameter constraints in section 5. Fig.~\ref{figA.2}
gives in the upper panel the effect of the cosmological parameters
and in the lower panel the effect of the scaling relation
parameters. We can see that e.g. an increase of the slope of the scaling
relation makes the XLF steeper, which can be compensated by a
larger  $\sigma_8$ counteracted by a smaller $\Omega_m$ to keep
the amplitude of the XLF in match.

\end{document}